\documentclass[useAMS,usenatbib]{mn2e}
 
\usepackage{amssymb,amsmath}
\usepackage{graphicx} 
\usepackage{xcolor}
\usepackage{float}
\usepackage{soul}

\begin{document}

\title[Radio Galaxy Shape Measurement with HMC]{Radio Galaxy Shape Measurement with Hamiltonian Monte Carlo in the Visibility Domain}
 
 \author[M. Rivi et al.]{
M. Rivi$^{1,2}$\thanks{E-mail: marzia.rivi@gmail.com},
M. Lochner$^{1,3,4}$,
S.T. Balan$^{1}$,
I. Harrison$^{5}$,
F.B. Abdalla$^{1,6}$
\\
$^{1}$Department of Physics and Astronomy, University College London, Gower Street, London, WC1E 6BT, UK\\
$^{2}$INAF - Istituto di Radioastronomia, via Gobetti 101, 40129 Bologna, Italy\\
$^{3}$African Institute for Mathematical Sciences, 6 Melrose Road, Muizenberg, 7945, South Africa\\
$^{4}$South African Radio Astronomy Observatory, The Park, Park Road, Pinelands, Cape Town 7405, South Africa\\
$^{5}$Jodrell Bank Centre for Astrophysics, School of Physics \& Astronomy, The University of Manchester, Manchester M13 9PL, UK\\
$^{6}$Department of Physics and Electronics, Rhodes University, PO Box 94, Grahamstown, 6140, South Africa
}

\maketitle

\begin{abstract}
Radio weak lensing, while a highly promising complementary probe to optical weak lensing, will require incredible precision in the measurement of galaxy shape parameters. In this paper, we extend the Bayesian Inference for Radio Observations model fitting approach to measure galaxy shapes directly from visibility data of radio continuum surveys, instead of from image data. 
We apply a Hamiltonian Monte Carlo (HMC) technique for sampling the posterior, which is more efficient than the standard Monte Carlo Markov Chain method when dealing with a large dimensional parameter space. Adopting the exponential profile for galaxy model fitting allows us to analytically calculate the likelihood gradient required by HMC, allowing a faster and more accurate sampling. The method is tested on SKA1-MID simulated observations at 1.4~GHz of a field containing up to 1000 star-forming galaxies. It is also applied to a simulated observation of the weak lensing precursor survey SuperCLASS. In both cases we obtain reliable measurements of the galaxies' ellipticity and size for all sources with SNR~$\ge 10$, and we also find relationships between the convergence properties of the HMC technique and some source parameters.
Direct shape measurement in the visibility domain achieves high accuracy at the expected source number densities of the current and next SKA precursor continuum surveys. 
The proposed method can be easily extended for the fitting of other galaxy and scientific parameters, as well as simultaneously marginalising over systematic and instrumental effects.
\end{abstract}


\section{Introduction}
\label{sec:intro}
The next generation of radio telescopes, such as the Square Kilometre Array (SKA)\footnote{https://www.skatelescope.org/}, will reach high enough sensitivity to provide a density of detected galaxies sufficient for weak lensing measurements in the radio band, with the advantage of reaching higher redshifts compared with optical surveys~\citep{Brown15, Harrison16}. Standard techniques for the measurement of cosmic shear are based on the observation of the shapes of faint star-forming (SF) galaxies, as it quantifies their coherent distortion by a large-scale foreground matter distribution. Such methods have been developed for optical surveys so far and require source shapes be measured accurately in order for errors to be dominated by statistics, rather than systematics \citep{Mandelbaum15}. Therefore they assume stringent and specific requirements on image fidelity. 
Radio instruments do not detect images of the observed sky, they provide its Fourier Transform at a finite number of points instead (\textit{visibilities}). The image reconstruction from these data, using iterative de-convolution methods such as CLEAN~\citep{clean74,clean78}, is a highly non-linear process that does not necessarily converge in a well-defined manner when dealing with extended sources resolved by high resolution telescopes. Moreover the Fourier Transform of visibilities makes the noise in radio images highly correlated. SKA simulations have shown that, even on high signal-to-noise ratio (SNR) objects, this process produces images with structures in the residuals that dominate the cosmological signal, confirming that this bias may be induced by the procedure adopted for turning the visibility data into images~\citep{Patel15}. 
Cross-correlation of real data images shows that there is no evidence of high levels of correlation between the optical and radio intrinsic shape of the matched objects~\citep{Patel10, THB16}, possibly due to systematics of the imaging pipeline. However a significant percentage of contaminating AGN sources (as they carry a much higher shape noise) and the astrophysical scatter between optical and radio position angle due to the different emission mechanisms may also be the reasons for such results. 
A more natural approach to adopt with radio data is to work directly in the visibility domain, where the noise is purely Gaussian and the data not yet affected by the systematics introduced by the imaging process, with the advantage of an exact modelling of the sampling function. However, this approach faces difficult statistical and computational challenges because sources are no longer localized in the Fourier domain and a telescope like SKA generates a large number of visibilities. Nevertheless, the advantages of being able to reduce, as well as holistically incorporate, systematic effects, coupled with Gaussian noise properties make modelling in the visibility domain a compelling option.

Bayesian Inference for Radio Observations (BIRO) \citep{BIRO} combines simulation tools, such as \textit{MeqTrees}~\citep{noordam}, capable of predicting visibilities from a given sky model and observational setup, with Bayesian inference samplers, like MCMC~\citep{metropolis,hastings} or Nested Sampling~\citep{skilling}. This allows full forward modelling of scientific properties, such as the flux and positions of sources, as well as instrumental effects, like pointing errors in telescopes, that can have strong effects on the scientific parameters. In addition, by making full use of the visibility information (instead of gridding as imagers do), BIRO can dramatically increase the resolving power of a telescope. BIRO could thus potentially be a useful tool for radio weak lensing, if the galaxies in the survey can be correctly modelled.
Other available tools for source fitting in the visibility domain have been developed for general purposes \citep{MVidal14}. They provide a variety of galaxy shape models, but they are not sufficiently accurate or capable on wide fields of large numbers of sources for weak lensing measurements. 

Currently, two analytical galaxy models have been considered for shear measurement in the visibility domain. 
The first one uses \textit{shapelets}~(\citealt{Refregier03}; \citealt{RB2003}), where galaxy images are decomposed through an orthonormal basis of functions corresponding to perturbations around a circular Gaussian. Shapelets are invariant under Fourier Transform (up to a rescaling) allowing the adoption in the visibility domain of the same approaches used with images. Finding the best fitting shape via minimising the chi-squared  is nominally linear in the coefficients and therefore can be performed simultaneously for all sources by simple matrix multiplications. However, this is only the case once a size scale $\beta$ and number of shapelet coefficients to include have been chosen, which can in itself be a highly non-linear and time-consuming problem.
\citet{CR02} successfully applied this technique to data from the Faint Images of the Radio Sky at Twenty cm (FIRST) survey~\citep{Becker95}, conducted with the NRAO Very Large Array (VLA). They were also able to detect cosmic shear with a significance of 3.6$\sigma$ after an accurate treatment of systematic effects~\citep{Chang04}. However shapelets introduce a shear bias as they cannot accurately model steep brightness profiles and highly elliptical galaxy shapes \citep{Melchior10}. 
For this reason, \textit{S\'{e}rsic} models are commonly used in optical weak lensing surveys analysis \citep{Mandelbaum15}, where the disc component is described by the exponential brightness profile (S\'{e}rsic index $n=1$) and the bulge is approximated by the deVacouleur's profile (S\'{e}rsic index $n=4$).  A reasonable assumption is to use a single optical disc-like component in the radio band because the radio-emitting plasma should follow the distribution of stars in galaxy discs, as it is due to the synchrotron radiation of the interstellar medium. The possible bias arising for this model has been discussed for the optical domain in \citet{VB10, Miller13, modelBias}. A similar model bias should be expected in the visibility domain.

In \cite{Rivi16} the exponential profile has been Fourier transformed analytically and used to extend the optical semi-Bayesian \textit{lensfit} method~\citep{Miller13} to radio data in the visibility domain.  
This method, called \textit{RadioLensfit}, fits the visibilities of a single source at a time and applies a Bayesian marginalisation of the likelihood over uninteresting parameters.  Visibilities of each galaxy are isolated by a source extraction algorithm described in \cite{Rivi18}. This approach is very fast computationally but may be limited by the source number density in the field of view, because of nearby galaxies residuals in the extraction procedure (``neighbour bias"). 

We investigate in this paper a different approach for galaxy shape measurement in Fourier space to perform a joint model fitting of all the sources in the field of view. We adopt the same analytical galaxy model used in \textit{RadioLensfit} and introduce it into the BIRO formalism. Unlike in \cite{BIRO}, we now have a large number of parameters to cope with so we use a Hamiltonian Monte Carlo (HMC) sampler \citep{Neal11}.  
We first test the method on two SKA1-MID simulated data, investigating also the potentiality of fitting a large number of sources. Then we apply the method on the simulation of a real observation: SuperCLASS (Battye et al, in prep), whose UV coverage is quite complex as it is composed of the baselines of two different SKA precursor radio telescopes (e-MERLIN and JVLA). Finally we discuss the convergence of individual parameters depending on source properties. 

Our paper is organised as follows. In Section~\ref{sec:method} we outline the method and its implementation. In Section~\ref{ska-sim} and~\ref{SC-sim} we describe data simulation and results for SKA1 and SuperCLASS respectively. In Section~\ref{convergence} we discuss HMC convergence. We also discuss how to handle AGN contamination in real observations in Section~\ref{AGN}. Finally, conclusions are presented in Sections~\ref{conclusions}.
 
\section{The Method}
\label{sec:method}
In this work we propose to measure radio galaxy shapes directly from the raw visibilities (i.e. in the Fourier space) by sampling the joint posterior distribution of the shape parameters of all the detected galaxies in the field of view. Since the number of galaxies in a single pointing may be very large, conventional Markov-Chain Monte Carlo (MCMC) methods using random-walk samplers (e.g. Metropolis-Hastings) are not suitable because they require a prohibitive number of samples to explore high-dimensional parameter spaces. We use the Hamiltonian Monte Carlo  approach that reduces random walk behaviour by applying the Hamiltonian dynamics of particles in potential wells (Section~\ref{sec:hmc}). It requires the computation of the likelihood gradient that we can compute analytically for our purpose, avoiding numerical differentiation (Section~\ref{sec:likelihood}). Mean and standard deviation of the sampled chain provides the measure and related error of the model parameters.

\subsection{HMC Algorithm}
\label{sec:hmc}
Hamiltonian Monte Carlo \citep{Neal11, Betancourt17} is a MCMC method that adopts physical system dynamics to explore a target distribution efficiently, resulting in faster convergence and maintaining a reasonable efficiency even for high dimensional problems~\citep{Hanson01,TAH08}.

The basic idea is to sample a distribution $P(\mathbf{x})$ of $n$ parameters $\mathbf{x}$ according to the Hamiltonian of a $2n$-dimensional dynamical system, where the parameters correspond to the positions of $n$ particles and $P(\mathbf{x})$ is the canonical distribution of the potential energy function $U(\mathbf{x})$. Momentum variables $\mathbf{p}$ are introduced to be coupled with positions $\mathbf{x}$ to allow Hamiltonian dynamics to operate according to a kinetic energy function $K(\mathbf{p})$ whose canonical distribution is the zero-mean Gaussian with a diagonal covariance matrix $\mathbfss{M} = \mathrm{diag}(m_1,\ldots,m_n)$, where $m_i$ represents the mass of particle $i$. This means:
\begin{equation}
U(\mathbf{x}) = -\log[P(\mathbf{x})], \qquad K(\mathbf{p})= \frac12 \mathbf{p}^T \mathbfss{M}^{-1} \mathbf{p}.
\end{equation}
Samples from the parameters distribution are obtained by marginalising the canonical distribution of the corresponding Hamiltonian function $H({\bf x},{\bf p}) = U({\bf x}) + K({\bf p})$:
\begin{equation}
\exp[-H(\mathbf{x},\mathbf{p})] = P(\mathbf{x})\exp[-\frac12 \mathbf{p}^T \mathbfss{M}^{-1} \mathbf{p}]
\end{equation}
over $\mathbf{p}$.

In practice, at each iteration of HMC a new sample $\mathbf{x}'$ is generated from the previous one $\mathbf{x}_0$ as follows. First a proposal $\mathbf{p}_0$ for the momentum variables is generated randomly according to the canonical distribution of the kinetic energy, i.e. a $n$-dimensional uncorrelated Gaussian. Then, a deterministic proposal for positions $\mathbf{x}$ is computed by allowing the system to evolve for a fixed time $\tau$ from the starting point $(\mathbf{x}_0, \mathbf{p}_0)$ according to Hamilton's equations:
\begin{align}
\frac{\mathrm{d} x_i}{\mathrm{d} t} & = \frac{\partial H}{\partial p_i} = [\mathbfss{M}^{-1} \mathbf{p}]_i\\
\frac{\mathrm{d} p_i}{\mathrm{d} t} & = - \frac{\partial H}{\partial x_i} = - \frac{\partial U({\bf x})}{\partial x_i}. 
\end{align} 
These equations are usually solved numerically by applying the \textit{leapfrog} method, where Hamilton's equations are discretized with a time step of small size $\varepsilon$ as follows:
\begin{align}
p_i \Bigl(t+\frac{\varepsilon}2 \Bigr) & = p_i(t) - \frac{\varepsilon}2 \frac{\partial U}{\partial x_i} \Bigr|_{{\bf x}(t)} \\
x_i(t+\varepsilon) & = x_i(t) + \frac{\varepsilon}{m_i} p_i \Bigl( t+\frac{\varepsilon}2 \Bigr) \\
p_i(t+ \varepsilon) & = p_i \Bigl( t+\frac{\varepsilon}2 \Bigr) - \frac{\varepsilon}2  \frac{\partial U}{\partial x_i} \Bigr|_{{\bf x}(t+\varepsilon)}.
\end{align}
The new proposal for $\bf x$ is obtained after $k$ steps as ${\bf x}'={\bf x}(\tau)$, with $\tau=k\varepsilon$. It is accepted with Metropolis criterion, i.e. with probability:
\begin{equation}
\min\{1,\exp(-\delta H)\}, 
\end{equation} 
where 
$$\delta H = H(\mathbf{x}',\mathbf{p}') - H(\mathbf{x}_0,\mathbf{p}_0).$$
This way the rate of acceptance is close to unity when numerical solutions of Hamilton's equations are very close to the exact ones because along such a trajectory energy is conserved. \cite{Neal11} show that $\varepsilon <  2$ must be required in order to have stable trajectories computed with the leapfrog method.
Our C++ implementation of the HMC algorithm will soon be available within a BIRO repository.

\subsection{Likelihood and gradient computation} 
\label{sec:likelihood}
In order to sample the posterior of the parameters, we need to define the likelihood function obtained by comparing data visibilities $\tilde V_i$ with the model visibilities $V_i$ dependent on parameters $\bf x$:
\begin{equation}
\mathcal{L}({\bf x}) = \left(\displaystyle\prod_i^N\frac{1}{\sqrt{2\pi\sigma_i^2}}\right)\exp \left[ - \sum_i \frac{|\tilde V_i - V_i({\bf x})|^2}{2\sigma_i^2}\right], 
\end{equation}
where $\sigma_i$ is the standard deviation of the Gaussian noise on the $i$-th data visibility.

Visibilities are evaluated at the interferometer baseline vectors, whose coordinates $(u,v,w)$ are measured in wavelengths at the centre frequency of the signal band with respect to a coordinate system where $w$ points towards the phase tracking centre~\citep{Thompson}.
For a sky containing $N$ galaxies with exponential profile, we adopt the \textit{RadioLensfit} analytical visibility model \citep{Rivi18}:
\begin{equation}
\label{model}
V(u,v,w) =  \Big( \frac{\lambda_\textrm{ref}}{\lambda}\Big)^\beta \sum_{s=1}^N  \frac{S_{s,\lambda_\textrm{ref}}\mathrm{e}^{2\pi \mathrm{i}\bigl(u l_s + v m_s+w \sqrt{1-l_s^2-m_s^2}\bigr)}}{\big(1+4\pi^2 \alpha^2_s |\mathbfss{A}_s^{-T}\mathbf{k}|^2\big)^{3/2}}
\end{equation}
where $\beta=-0.7$ is the spectral index for the synchrotron radiation emitted by the galaxy disc, $\mathbf{k}=(u,v)$, and for $s=1,\ldots,N$ we have the following source $s$ parameters: position coordinates $(l_s,m_s)$, flux~$S_{s,\lambda_\textrm{ref}}$ at reference wavelength $\lambda_\textrm{ref}$, scalelength~$\alpha_s$ and ellipticity components $e_{1,s}, e_{2,s}$. The ellipticity parameters are contained in the matrix~$\mathbfss{A}_s$ that transforms the circular exponential profile\footnote{$I(r) = I_0 \exp(-r/\alpha)$} in elliptical
\begin{equation}\label{linearTransf}
\mathbfss{A}_s  = \left( \begin{array}{cc} 1-e_{1,s} & -e_{2,s} \\ -e_{2,s} & 1+e_{1,s} \end{array} \right).
\end{equation}
We assume the following ellipticity definition:
\begin{equation}
\mathbf{e} = e_1 +\mathrm{i}e_2 = \frac{a-b}{a+b}\mathrm{e}^{2\mathrm{i}\theta},
\end{equation}
where $a$ and $b$ are the galaxy major and minor axes respectively, and $\theta$ is the galaxy orientation.
Note that, in case future radio observations show that a 2-component galaxy model is also required in the radio domain, a similar analytical model for the bulge in the Fourier domain approximating the deVacouleur's profile is provided in \citet{Spergel10}.

A preliminary data analysis for measuring flux and position of the each source may allow the reduction of the number of parameters to $3N$. This can be achieved either in the image domain\footnote{Based on SKA1 level 0 science requirements, SKA1-MID shall provide astrometric accuracy of at least 1\%.} or, if necessary, in the visibility domain by a multimodal nested sampling approach adopting a single source model \citep{GalNest}. On the other hand, including these parameters in the full HMC analysis will require a tight prior on the positions in order to have a small additional computational complexity. In this work we assume them well known.  

Since the visibility model is defined analytically we can compute the exact likelihood gradient with respect to the parameters of our interest: scalelength and ellipticity components of each source in the primary beam.
The computation of the galaxy exponential model and the corresponding likelihood gradient (see Appendix~\ref{appendix}) has been added to the open-source software Montblanc\footnote{https://github.com/ska-sa/montblanc/tree/chi\_sqrd\_gradient}
 \citep{Montblanc}, which is a GPU accelerated implementation of the Radio Interferometric Measurement Equation (RIME)~\citep{Hamaker96, SmirnovA, SmirnovB} in support of BIRO. The RIME is a powerful framework to easily describe what happens to a signal as it travels from source to the interferometer in terms of Jones matrices, where multiple effects along the signal propagation path correspond to multiplication of matrices describing each effect. For example, equation~(\ref{model}) can be re-arranged as follows:
\begin{equation}
\label{eq:vis}
V_{tpq\lambda} = \sum_{s=1}^N V_{tpq\lambda s} = \sum_{s=1}^N  K_{tps\lambda} B_{s\lambda} K^H_{tqs\lambda}  
\end{equation}
where uvw points are identified by the baseline formed by antennas $p$ and $q$ at timestep $t$ and channel with centre wavelength $\lambda$. $K_{tps\lambda}$ and $B_{s\lambda}$ are respectively the phase and brightness matrices for source $s$ ($H$ denotes matrix hermitian transpose). We use a brightness matrix defined by the exponential model of an extended source\footnote{Other galaxy models available in Montblanc are Gaussian and point sources. For Gaussians a simpler computation of the likelihood gradient can be added.}:
\begin{equation}
\label{eq:brightness}
B_{s\lambda} = \Big( \frac{\lambda_\textrm{ref}}{\lambda}\Big)^\beta \frac{S_{s,\lambda_\textrm{ref}}}{\big(1+4\pi^2 \alpha^2_s |\mathbfss{A}_s^{-T}\mathbf{k}|^2\big)^{3/2}}.
\end{equation}
Given the general formalism of the RIME, any instrumental effects that can be modelled, such as primary beam shape, pointing errors and instrumental noise, can be incorporated and marginalised over with our formalism. 

\subsection{Priors}
\label{sec:priors}
As source scalelength and ellipticity priors we use the distributions presented in \cite{Rivi18},  whose parameters are estimated from VLA observations:
\begin{itemize}
\item $p(\alpha)$ is a lognormal function with  mean $\mu = \ln(\alpha_\mathrm{med})$ and variance $\sigma \sim 0.3$, where $\alpha_\mathrm{med}$ is given by equation~(\ref{scale-flux});  
\item for the modulus $e$ of the intrinsic galaxy ellipticity  
\begin{equation}
p(e) = \frac{Ne\left(1-\exp\left[ \frac{e-e_\textrm{max}}{c}\right]\right)}{(1+e)(e^2+e_0^2)^{1/2}}, 
\end{equation}
with $c =  0.2298$, $e_0 = 0.0732$ and $e_\mathrm{max} = 0.804$.
\end{itemize}
These distributions are used for both the galaxy catalog simulation and the prior in the HMC sampling.
  
\subsection{HMC tuning}
\label{sec:code}
\begin{figure*}
\includegraphics[scale=0.285]{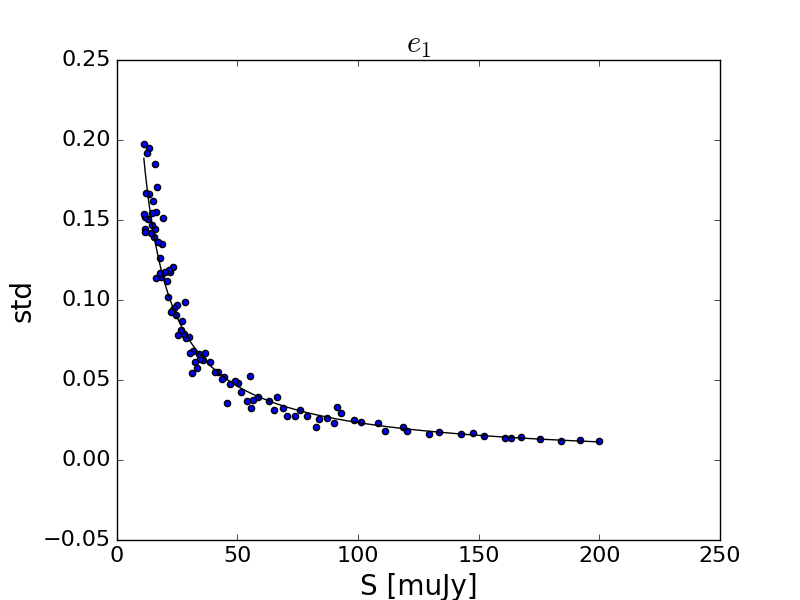}
\includegraphics[scale=0.285]{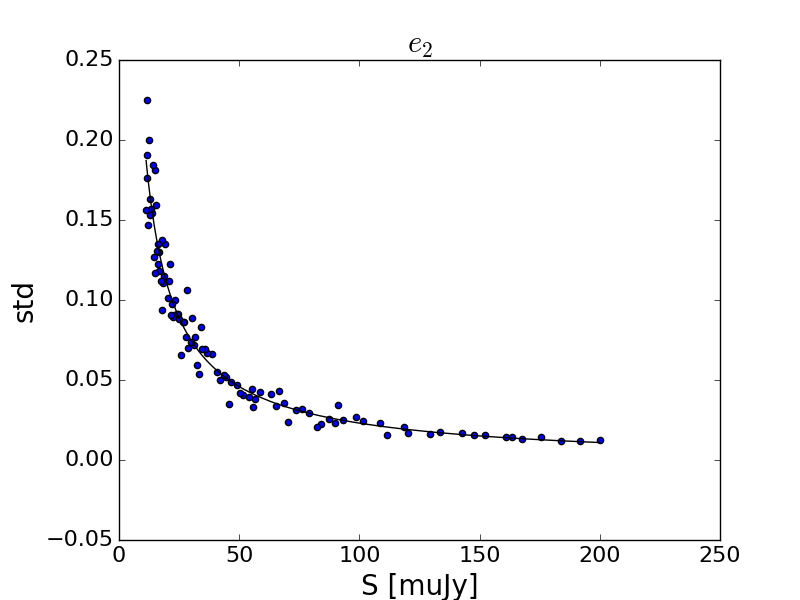}
\includegraphics[scale=0.285]{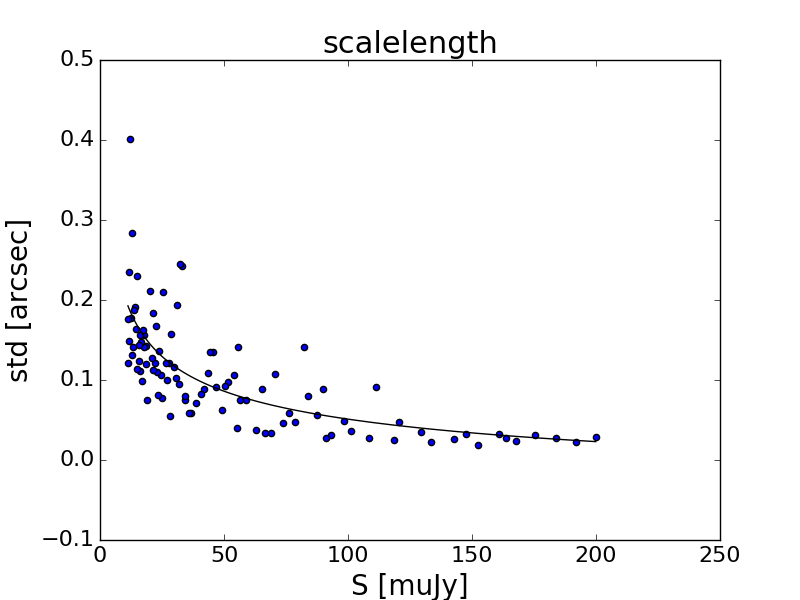}
\caption{Power law function fitting the standard deviation of the parameters versus integrated source flux from the SKA1-MID measurements of 100 sources with SNR$_\mathrm{vis} \ge 10$.}
\label{fig:std}
\end{figure*}

Selecting suitable values of the step size $\varepsilon$ and the number of steps $k$ for the leapfrog method is crucial as discussed in \cite{Neal11}. After preliminary runs, we choose $\varepsilon=0.05$ and $k=10$ as a trade-off between the accuracy of the Hamilton's equations solutions (i.e. high acceptance rate) and the computational time due to a too long trajectory.  

Since most of galaxy ellipticities should be close to zero, according to the ellipticity modulus distribution, we choose as starting points for the ellipticity components random numbers uniformly distributed  between $-0.1$ and $0.1$.
For scalelength parameters we choose as starting points the median values depending on the source flux density according to the relation estimated in \cite{Rivi15} from the VLA 20~cm continuum radio source catalog in the SWIRE\footnote{http://heasarc.gsfc.nasa.gov/W3Browse/radio-catalog/vlasdf20cm.html} field~\citep{OM2008}:
\begin{equation}
\label{scale-flux}
\ln{[\alpha_\mathrm{med}/\textrm{arcsec}]} = -0.93 +0.33\ln{[S/\mu \textrm{Jy}]}.
\end{equation}
For the convergence of HMC chains, it is very important to choose good ``step sizes'' of the momentum samples. They are defined by the inverse mass in the kinetic energy function and typically are given by the variance of the parameters. In this application parameter variance is dependent on the source signal-to-noise ratio, and therefore integrated source flux. We estimate a relation between the standard deviation for source scalelength and ellipticity components as a function of its flux density by fitting the error bars measured from a simulated observation of 100 sources (as described in Section~\ref{ska-sim}).  
For both ellipticity components and scalelength we obtain a power law relation between source flux (in $\mu$Jy) and measured standard deviation. For example, for SKA1-MID observations we obtain the following relations (see Fig.~\ref{fig:std}):
\begin{align}
& \sigma_{e_1}, \sigma_{e_2} =  -0.0022 + 1.77 S^{-0.92}, \\
& \sigma_{\alpha}[\mathrm{arcsec}] =  -0.0914 + 0.61 S^{-0.31}.
\end{align}

\section{SKA1-MID simulations}
\label{ska-sim}
\begin{figure}
 \centering
  \includegraphics[width=\columnwidth]{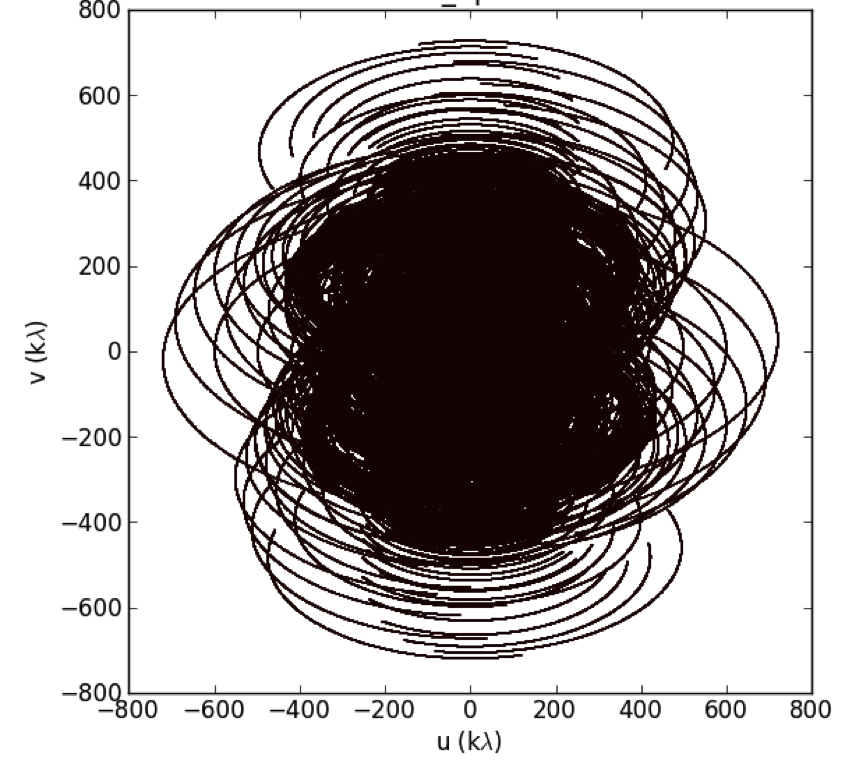} 
  \caption{UV coverage of SKA1 observations.}
  \label{fig:ska1}
\end{figure}
We simulate SKA1-MID 8-h observations pointing close to the zenith ($\delta \sim -30^\circ$). Visibilities are sampled every 60~sec at the frequency 1.4~GHz (the corresponding UV coverage is plotted in Fig.\ref{fig:ska1}) and we choose a single large channel with bandwidth 240~MHz, as no smearing effects are added.    
Visibilities are simulated by using equation~(\ref{model}) and adding an uncorrelated Gaussian noise whose variance is dependent on the antenna system equivalent flux density (SEFD) of SKA1 dishes, the frequency channel bandwidth and the time sampling according to the formula given in \cite{Sensitivity99}. 

\subsection{Source catalogs}
The source catalogs are simulated as in \cite{Rivi18}, where positions are generated randomly according to a uniform distribution over a circular field of view, the flux distribution is $p(S) \propto S^{-1.34}$, scalelength and ellipticity follow the prior distributions listed in Section~\ref{sec:priors}. 

The minimum integrated source flux detectable at a SNR$_\mathrm{vis}$\footnote{We compute the signal-to-noise ratio in the visibility domain as SNR$_\mathrm{vis} = \sqrt{\sum_{i=1}^\textrm{nvis} |V_i|^2/\sigma^2_i}$, where $V_i$ are the simulated visibilities without noise.}~$=10$ is 10~$\mu$Jy, and the source density expected in the current planned SKA1 radio weak lensing survey (2.7 gal/arcmin$^2$, \cite{Brown15}) is obtained at the maximum flux of~200~$\mu$Jy. We choose a field of view of 400~arcmin$^2$ to test the method at such density for 1000 galaxies (``SKA1~1000''). To compare measurement accuracy and the speed of convergence as a function of the number of sources to fit, we also test the case of 100 sources with the same flux range and field of view (``SKA1~100").

\subsection{Results}

\begin{figure*}
 \centering
 \includegraphics[width=1.8\columnwidth]{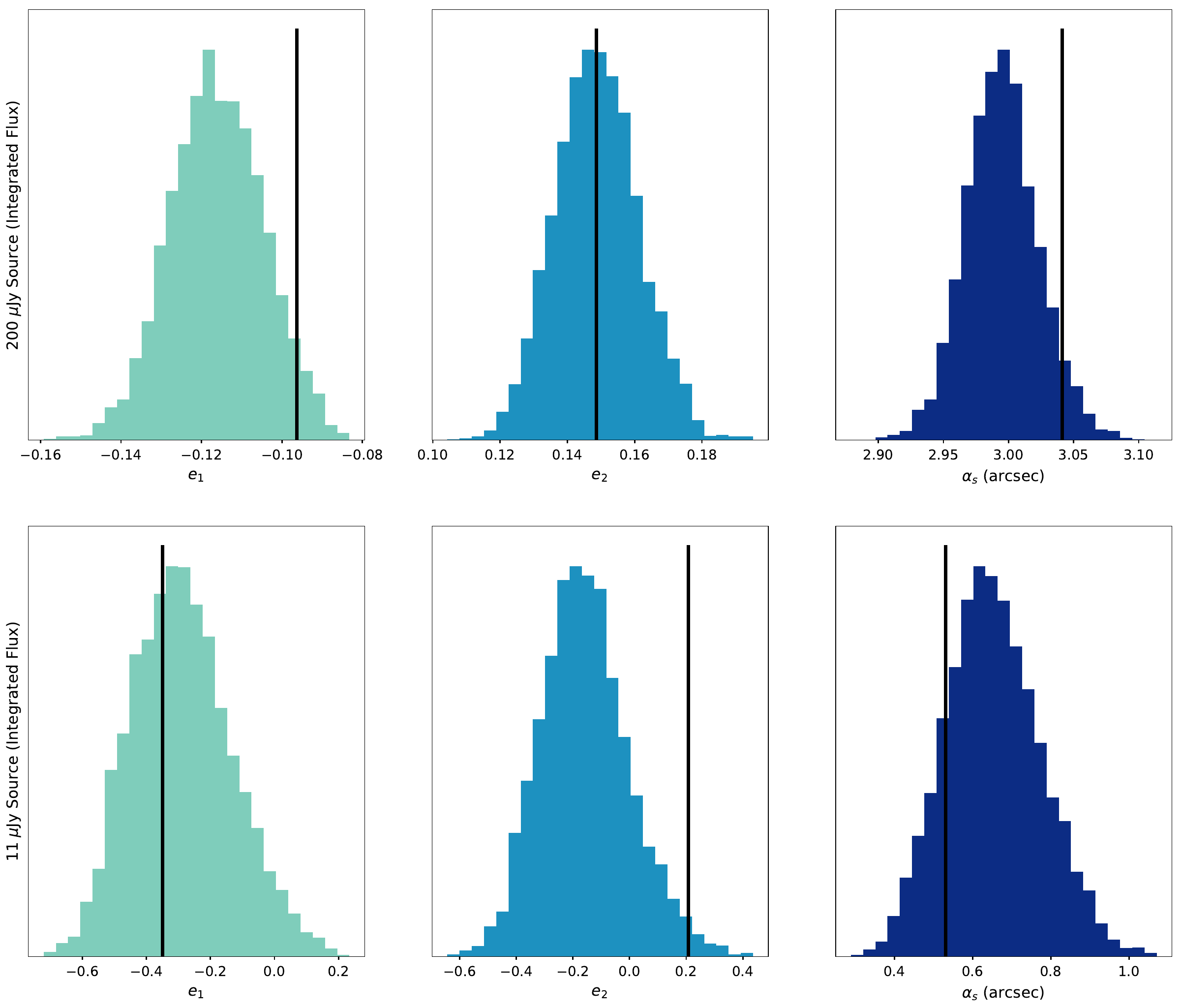}
 \caption{Example 1-d marginalised posterior plots for the SKA1 100 source case, for the brightest (top panel) and the faintest (bottom panel) 
sources in the catalog. The input value for each parameter is shown by a solid line, illustrating the ability of our approach to accurately recover both the parameter value and its uncertainty.}
\label{fig:hists}
\end{figure*}

\begin{figure*}
 \centering
 \begin{minipage}{2.1\columnwidth}
 \begin{minipage}{0.99\linewidth}
 \includegraphics[width=0.99\linewidth]{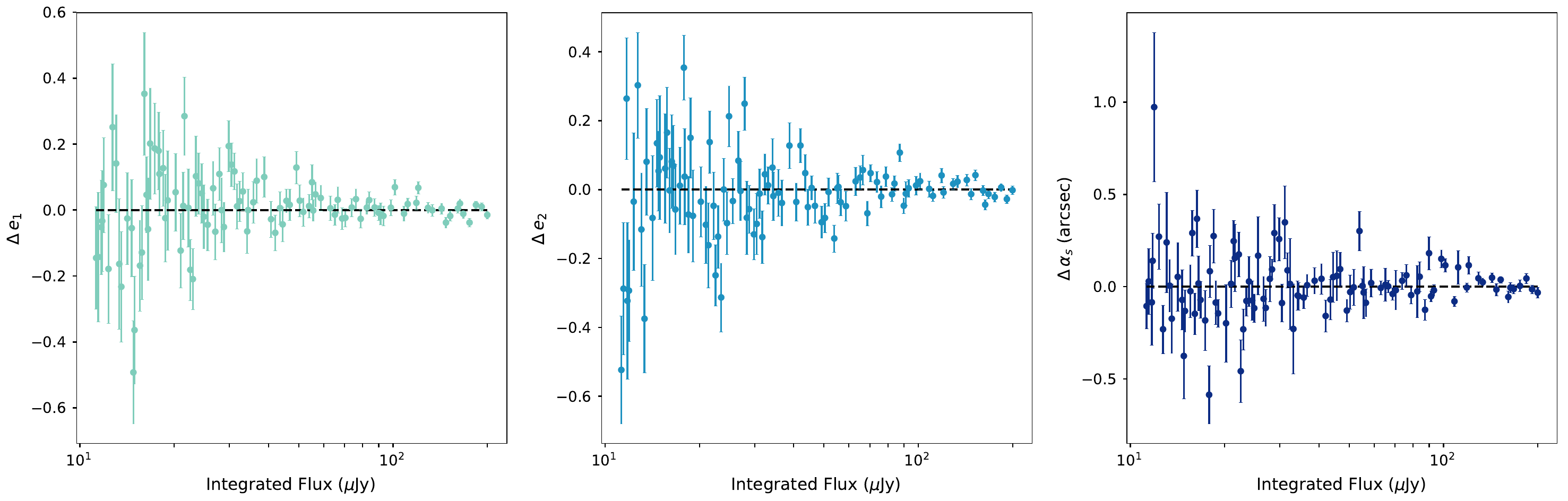} 
 \end{minipage}
 \begin{minipage}{0.99\linewidth}
 \includegraphics[width=0.99\linewidth]{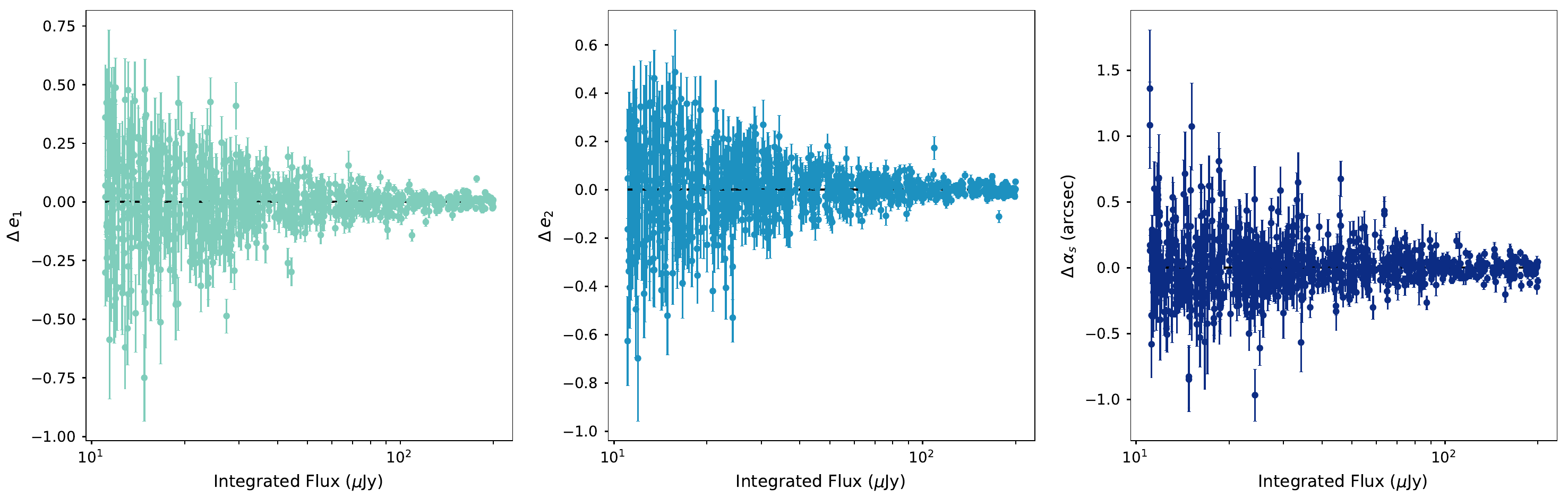} 
 \end{minipage}
 \end{minipage}
 \caption{The measured minus true value of each parameter as a function of integrated flux for the ellipticity parameters ($e_1$ and 
$e_2$) and the scalelength parameter ($\alpha_s$), for SKA1 100 sources (top panel) and SKA1 1000 sources (bottom panel). The error bar is estimated from the 1-d marginalised posterior for that parameter. It is clear that brighter sources have lower uncertainty on the parameters.}
\label{fig:results}
\end{figure*}

In Fig.~\ref{fig:hists} we show the one-dimensional marginalised posterior for a few example parameters of ``SKA1-100" showing they are well-constrained and accurately recover input parameters.
We show the difference between the measured and input parameters (which should be close to zero) for both SKA1 cases in Fig.~\ref{fig:results}. As expected, brighter objects have much lower uncertainty in the parameters. We find that the shape parameters are largely uncorrelated although they could be highly correlated with instrumental effects that we do not model here \citep{BIRO}. 

Table~\ref{tab:SKAbest-fit} contains the values of the best-fit lines coefficients for the scalelength and the ellipticity components obtained by fitting the simulated observations of 100 and 1000 sources in the field of view respectively.  They are almost the same for both test cases, where only the uncertainty is obviously larger for a smaller number of sources. The goodness of fit for all shape parameters, estimated as a straight line fit of measured versus true parameters, shows that the measurements are very precise.
 
Comparing the best-fit slope of ellipticities for 1000 sources, i.e. at the source density of the proposed SKA1 weak lensing survey, with the one obtained at the same density with \textit{RadioLensfit} we see a significant improvement. In fact, in \cite{Rivi18} the best-fit slopes of the ellipticity components, measured from a similar source population and SKA1 UV coverage, were found to be $a_1=0.9365 \pm 0.0017$ and $a_2=0.9262 \pm 0.0017$ respectively. This improvement is expected because joint fitting avoids the source extraction bias of the \textit{RadioLensfit} method, but with a large computational cost (see Table~\ref{tab:convergence}). 
We also observe that ellipticity error bars of our approach are larger, but correctly measured, mainly for two reasons: (i) \textit{RadioLensfit} may underestimate the parameters uncertainty by sampling the likelihood only in a neighbourhood of the maximum point (see Section~3.3 of \cite{Rivi16}), while we marginalise the posterior instead; (ii) in this work we have a larger number of free parameters as we are measuring both ellipticity and scale-length of all sources simultaneously. Thus the larger error bars are actually an indication of the technique correctly incorporating additional sources of uncertainty.

\begin{table*}
\begin{tabular}{l c c c c c c}
\hline
\textbf{Observation} & \multicolumn{2}{|c|}{$\mathbf{e_1}$} & \multicolumn{2}{|c|}{$\mathbf{e_2}$} & \multicolumn{2}{|c|}{\textbf{scalelength}}  \\
  & $a$ & $c$ & $a$ & $c$  & $a$ & $c$ \\
\hline
SKA1-100 &  $0.9559 \pm 0.0116$ & $-0.0018 \pm 0.0032$ & $ 0.9780 \pm 0.0118$ & $ -0.0014 \pm 0.0031$ & $1.0015 \pm 0.0091$ & $-0.0019 \pm 0.0161$ \\
SKA1-1000 & $0.9704\pm 0.0043$ & $0.0001 \pm 0.0010$ & $0.9718 \pm 0.0040$ & $-0.0002 \pm 0.0010$ & $1.0048 \pm 0.0030$ & $-0.0090 \pm 0.0051$ \\
\hline
\end{tabular}
\caption{Multiplicative (\textit{a}) and additive (\textit{c}) coefficients of the best-fit lines for galaxy shape measurements with SKA1 at SNR$_\mathrm{vis} \ge 10$.} 
\label{tab:SKAbest-fit}
\end{table*}

\section{Application to SuperCLASS}
\label{SC-sim}
We also investigate the ability of the method to infer shear using current-generation data, from the SuperCLASS survey. This allows us to both assess the possibilities in near-term data, and also understand the effect of realistic UV coverages, which are not typically observed as a single long track, rather multiple interleaved pointings. We emphasise that the results we generate here should not be used as a measure of the relative shape measurement ability of SuperCLASS and an SKA survey; shape measurement ability will be a strong function of UV coverage and we use a very realistic coverage for SuperCLASS and a highly simplified (and much smaller in volume) coverage for SKA.

\subsection{The survey}
The SuperCLASS (Super CLuster Assisted Shear Survey)\footnote{http://www.e-merlin.ac.uk/legacy/projects/superclass.html} (Battye et al, in prep.) is a legacy survey on the UK's e-MERLIN radio telescope, with the express goal of making a convincing detection of a weak lensing signal in radio data. For the field, containing four massive galaxy clusters at $z \sim 0.2$, observations are taken at L-Band (around 1.4 GHz) using both e-MERLIN and the JVLA. This gives coverage of a wide range of spatial scales, with sensitivity to angular scales between 1 and 10 arcseconds principally coming from the JVLA, and smaller scales from the longer baselines available to e-MERLIN. The field has also been observed at multiple lower radio frequencies (to assist in Rotation Measure synthesis and source classification), sub-mm, and optical and near-IR wavelengths (to obtain photometric redshifts and optical shear measurements). The full data release of the survey is expected to contain thousands of galaxies at a density of $\sim 1\,$gal/arcmin$^2$ over $1\,$deg$^2$. Shape measurement for the survey is being performed using both calibrated image-plane and hybrid image/visibility-plane techniques.

\subsection{Data simulation}
We create simulated UV coverages which closely match the true ones in the SuperCLASS survey. This provides us with an idealised but realistic version of the experiment; though we do not include effects from calibration errors and missing data from telescope outages or Radio Frequency Interference \citep[RFI, an appreciable problem for e-MERLIN,][]{2013A&C.....2...54P}, we do recreate the real shape of the UV coverage (corresponding to the shape of the PSF). The UV coverages for both telescopes are simulated as measurement sets using the CASA simulator tool \citep{2007ASPC..376..127M}.

\subsubsection{e-MERLIN}
For e-MERLIN, we simulate 8 IFs of 512 channels of width 125 kHz, covering a bandwidth of 512 MHz upwards from a starting frequency of 1.25 GHz. The UV coverage is then generated as it is for a real observation: during a single eight hour observing run (one epoch), seven different pointing centres are observed, along with amplitude and phase calibration sources. Observations cycle around the seven pointing centres, spending alternately 12.5 minutes observing the source and 2.5 minutes observing calibrators. An individual pointing centre is observed over four epochs, giving a total amount of time on source of just over eight hours per pointing centre. This gives an image plane noise level of $\sim 14 \,\mu$Jy$/$beam before different pointings are mosaiced together; a process which reduces the noise by a factor of two.
The UV coverage resulting from this procedure is shown in the left panel of Fig.~\ref{fig:uv-coverage}.

\subsubsection{JVLA}
\begin{figure*}
\includegraphics[width=0.46\textwidth]{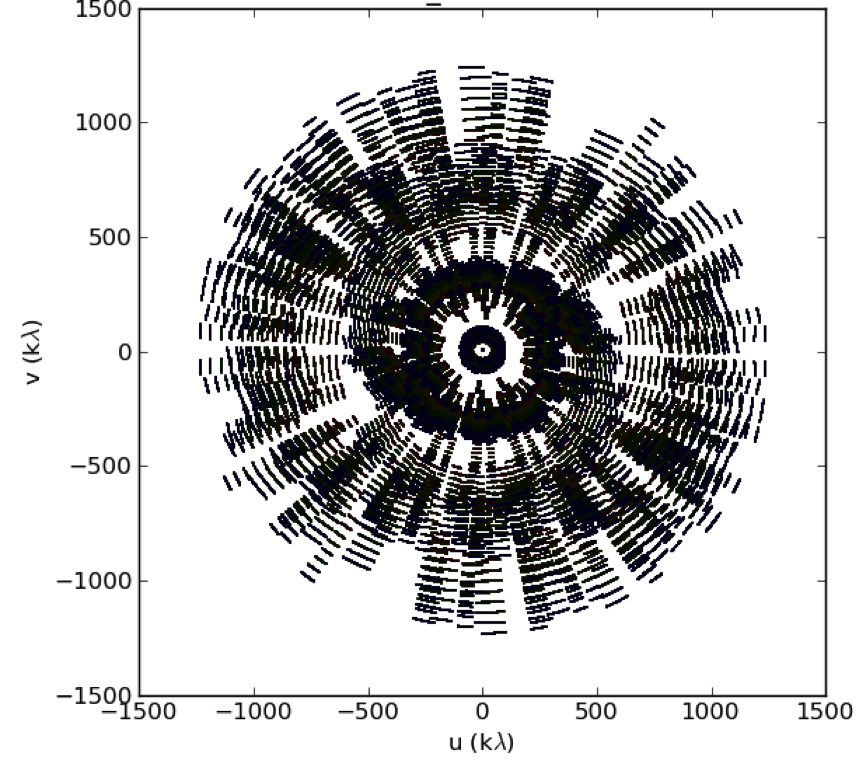}  
\includegraphics[width=0.45\textwidth]{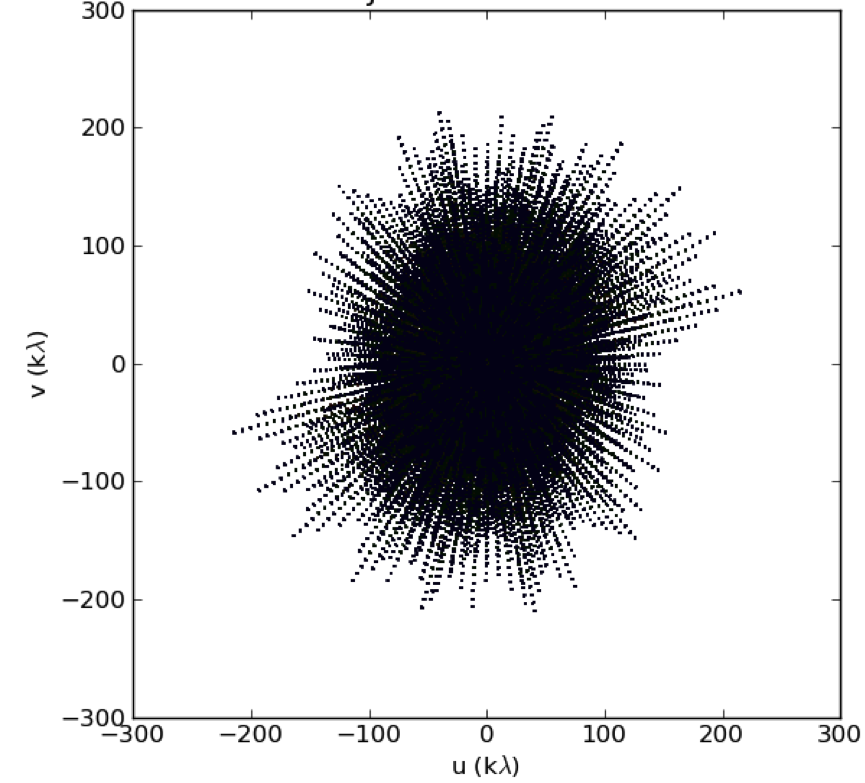} 
\caption{SuperCLASS UV coverage. Left: e-MERLIN. Right: JVLA (note the difference in scales).}
\label{fig:uv-coverage}
\end{figure*}

For the Jansky Very Large Array (JVLA) telescope observations of the SuperCLASS field, we simulate the full sixteen IFs of 256 channels each, with channel widths of 250 kHz, covering a bandwidth of 1024 MHz upwards from a starting frequency of 1.01 GHz. For the JVLA, each individual pointing is observed over six epochs, with a total time of 11 minutes 6 seconds on source. The greater filling of the Fourier plane on the sample scales by the JVLA when compared to e-MERLIN means that a similar image-plane depth of $\sim 14 \,\mu$Jy$/$beam is reached in this time. The UV coverage resulting from this procedure is shown in the right panel of Fig.~\ref{fig:uv-coverage}.

\subsubsection{Visibilities}
Because of restrictions on available memory we sample visibilities every 10 sec instead of the real sampling time of 1~sec. No time or bandwidth smearing effects are accounted for in the forward modelling. As in Section~\ref{ska-sim}, we simulate visibilities observed in a single pointing according to equation~(\ref{model}) and add uncorrelated Gaussian noise with variance dependent on the telescope configuration. In order to have the source SNR distribution expected for the real data, we artificially decrease this noise by a factor of two (as the real data will benefit from the mosaicing effect of overlapping pointings), considering only the pointing inner region of 100 arcmin$^2$ as field of view. The source number density of 1~gal/arcmin$^2$ at 10$\sigma$ in the image domain\footnote{Usually the source signal-to-noise in the image domain is given by the ratio between the source peak intensity and the image noise standard deviation. Since we are dealing with extended sources, this SNR tends to be lower than what measured in the visibility domain.} is obtained for a minimum input integrated source flux $S_\mathrm{min} \sim 500 \mu$Jy and, according to our flux prior, a maximum flux density of 400~mJy.
Note that the corresponding minimum source SNR in the visibility domain is about~$18$.    

\subsection{Shape measurements}
\begin{figure*}
 \begin{minipage}{0.99\linewidth}
 \includegraphics[width=0.99\linewidth]{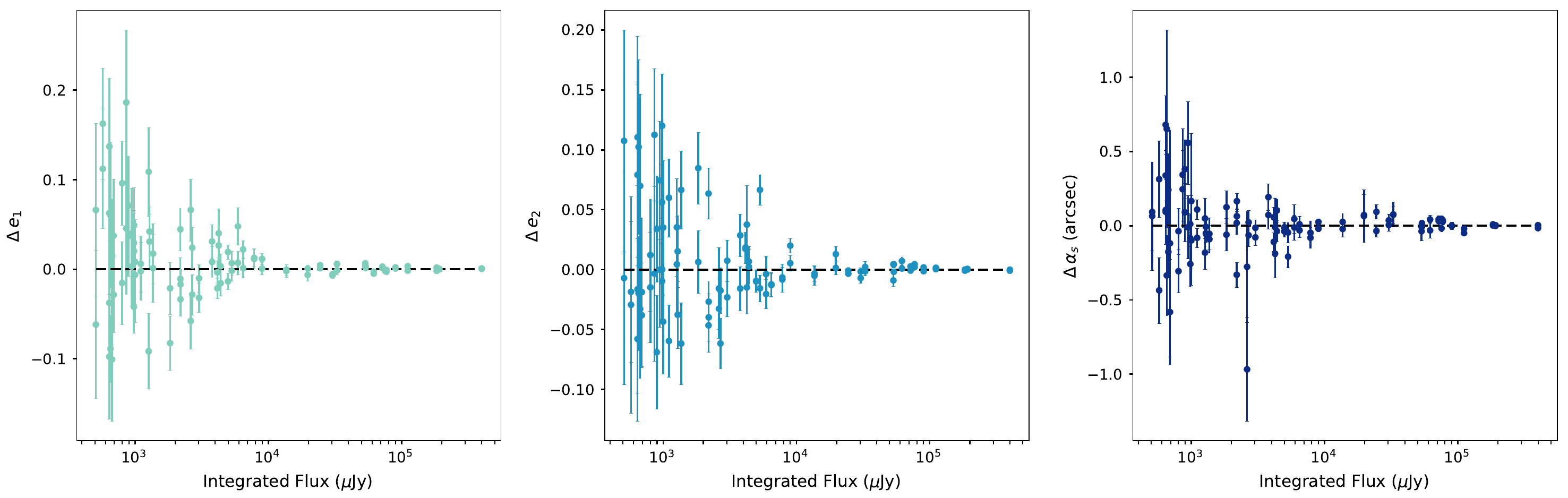} 
 \end{minipage} 
\caption{The measured minus true value of each parameter as a function of integrated flux for the ellipticity parameters ($e_1$ and $e_2$) and the scalelength parameter ($\alpha_s$) for SuperCLASS with 100 sources.}
\label{fig:SC-100}
\end{figure*}

\begin{table*}
\begin{tabular}{l c c c c c c}
\hline
\textbf{Observation} & \multicolumn{2}{|c|}{$\mathbf{e_1}$} & \multicolumn{2}{|c|}{$\mathbf{e_2}$} & \multicolumn{2}{|c|}{\textbf{scalelength}}  \\
  & $a$ & $c$ & $a$ & $c$  & $a$ & $c$ \\
\hline
SuperCLASS-100 & $1.0001 \pm 0.0010$ & $0.0006\pm 0.0002$ & $1.0011 \pm 0.0020$ & $0.0004 \pm 0.0002$ & $1.0002 \pm 0.0005$ & $-0.0047 \pm 0.0079$ \\
\hline
\end{tabular}
\caption{Multiplicative (\textit{a}) and additive (\textit{c}) coefficients of the best-fit lines for galaxy shape measurements with SuperCLASS at SNR$_\mathrm{vis} \ge 18$ (the lowest SNR in the SuperCLASS sample).} 
\label{tab:SCbest-fit}
\end{table*}
 
We are able to recover with a good accuracy the shape parameters of all sources. Fig.~\ref{fig:SC-100} shows the difference between the measured and input galaxy shape parameters, and Table~\ref{tab:SCbest-fit} contains the best-fit lines coefficients of the measured parameter values. The goodness of fit is better than for SKA1 simulations mainly because in this case the signal-to-noise ratio is higher (SNR$_\mathrm{vis} \ge 18$), but also because the UV coverage is much larger (see Table~\ref{tab:convergence}).

\section{Convergence analysis}
\label{convergence}

\subsection{Tests for convergence}
Any Bayesian inference sampler, such as HMC, is only guaranteed to converge to the true posterior with an infinite number of steps. Given that we obviously can only run a chain for a finite time, convergence must be tested for. We test for convergence of the HMC chains using two different metrics: the commonly-used Gelman-Rubin (GR) statistic 
\citep{gelman1992} and the autocorrelation function \citep{sokal1997}. 

\subsubsection{The Gelman-Rubin Statistic}
The GR is a measure of how similar a set of HMC chains are:  it compares the variance within a chain to the variance between chains and should be close to one for a 
converged set of chains. Ideally, one would run several independent chains, starting from different starting points. 
However, as we are computation limited in this work, we settle for running one chain that we split up into 
three separate chains and compare.  

For the SKA1 100 source chain and the SuperCLASS 100 source chain, we find GR $\approx 1.1$ for all parameters. However, for the SKA1 1000 source chain, the parameters converge more slowly and for a few parameters, GR $\approx 2$. While we still make use of this chain in our analysis, because we observe no strong bias with respect to the true parameters and find the effective sample size to be sufficient (see below), if used on real data a longer chain should ideally be run and convergence ensured.  

\subsubsection{Auto-correlation function}

\begin{figure}
 \centering
 \includegraphics[width=1.\columnwidth]{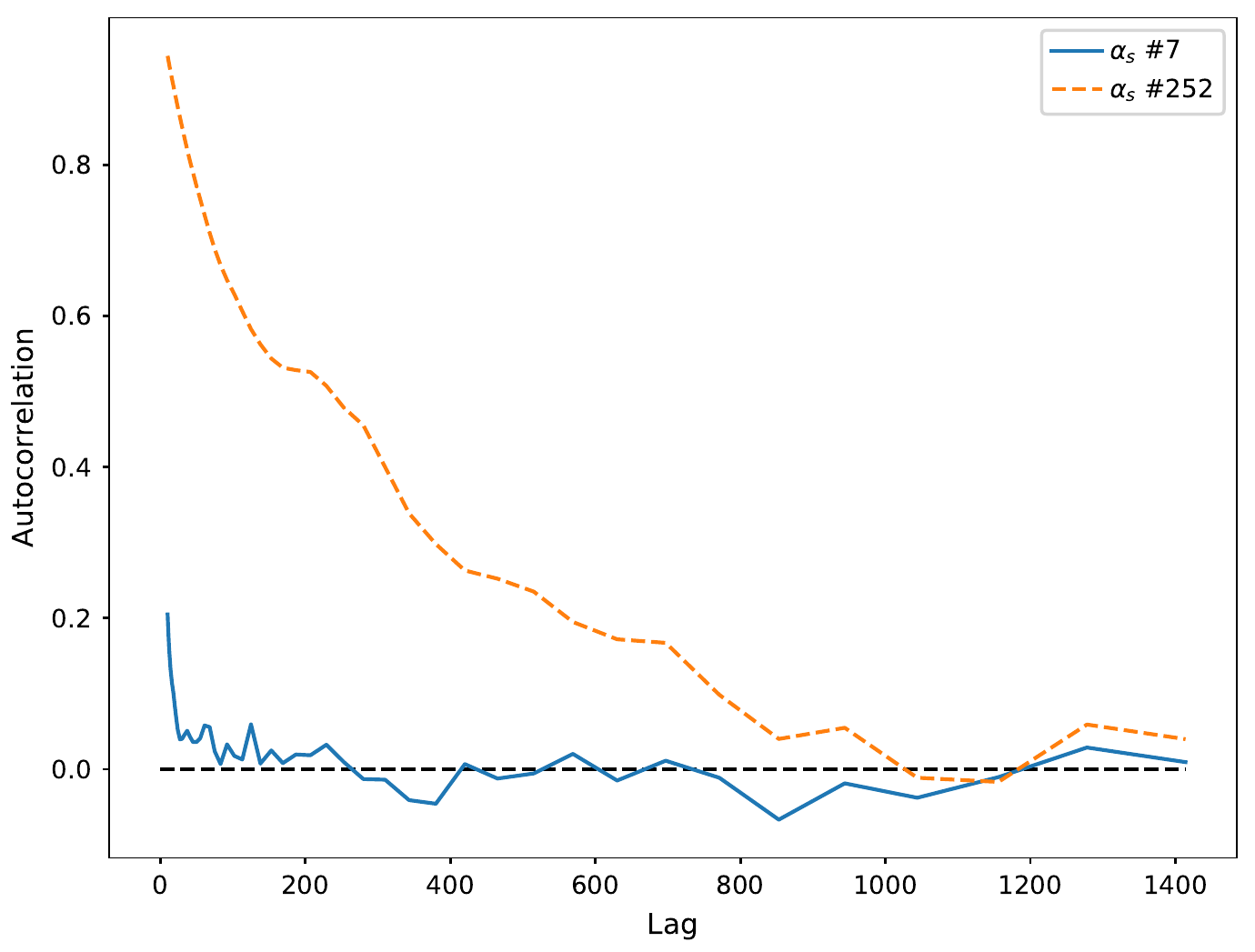}
 \caption{Autocorrelation curves for two example parameters of the SKA1 100 sources chain. When the 
autocorrelation is close to zero, steps separated by the corresponding lag are uncorrelated, independent samples. A 
short autocorrelation length corresponds to a well-mixed chain. Plotted here are the autocorrelation functions for the scalelength of a bright, round source (solid, blue line) and faint, highly elliptical source (dashed, orange line), for which the steps of the chain are still highly correlated indicating a difficulty to converge.}
\label{fig:correlation}
\end{figure}

An additional test for convergence is the lag-$k$ autocorrelation function which computes the autocorrelation between 
posterior samples separated by $k$ steps. The autocorrelation function for parameter $\theta$ , with mean $\bar \theta$, sampled by a length-$N$ chain is given by: 
\begin{equation}
\rho_k = \frac{\sum_i^{N-k}{(\theta_i-\bar\theta)(\theta_{i+k}-\bar\theta)}}{\sum_i^{N-k}(\theta_i-\bar\theta)^2}.
\label{eq:autocorrelation}
\end{equation}
$\rho_k=0$ indicates samples separated by $k$ are independent. 
Fig.~\ref{fig:correlation} shows examples lag-$k$ autocorrelation functions for some parameters of the ``SKA1-100'' chain. 
Samples in a chain from HMC are in general, not statistically independent.  The autocorrelation function is a measure of over what ``time scale'' steps are correlated. The faster the autocorrelation function drops to zero, the more independent samples can be drawn and the shorter the chain can be to achieve precision on parameter estimates. For a chain of a given length, this has direct impact on the error with which a parameter is measured.  

If the uncertainty on a given parameter as measured from the posterior is $\sigma$, then the \emph{sampling uncertainty} $\sigma_{\rm{samp}}$ is given by \citep{sokal1997}:
\begin{equation}
\sigma_{\rm{samp}}^2 = \frac{\sigma^2 t_{\rm{int}} }{N},
\label{eq:eff}
\end{equation}
where $N$ is the length of the chain. $t_{\rm int}$ is the \emph{integrated autocorrelation time} and is defined as:
\begin{equation}
t_{\rm int} = 1 + 2\displaystyle \sum_{k=0}^N \hat{\rho_k},
\label{eq:tint}
\end{equation}
where $\hat{\rho_k}$ is the normalised autocorrelation function $\rho_k/\rho_0$. $\sigma_{\rm samp}^2$ is the variance on the estimate of a given parameter which is due to finite sampling and is also sometimes referred to as the Monte Carlo standard error.

The integrated autcorrelation time is thus a key metric for analysing the convergence of a chain in an intuitive way: $t_{\rm int}$ dictates the number of steps required before a chain will reach a given level of precision in the estimate of a parameter, since it defines the number of independent samples that there are in a chain.  For instance, an \emph{effective sample size} ($N/t_{\rm int}$)  of  100 would contribute an additional $\sim 10\%$ to the uncertainty on a given parameter. As the effective sample size increases, the contribution to the error budget from sampling becomes negligible when compared with the intrinsic posterior uncertainty. 

In an ideal world, an HMC chain would always be run long enough for the sampling uncertainty to be too small to be concerned with. However, in reality computational resources are limited and the sampling uncertainty must be taken into account. The required level of precision depends on the final analysis to be done with the sample.  MCMC analyses have been done in the context of shear estimation in the case of the Dark Energy Survey \citep{2016MNRAS.460.2245J} where chains of a few thousand steps were found to be sufficient. Estimates of the required sampling uncertainty needed would critically depend on the design and cosmological requirements of that survey, which is beyond the scope of this paper. 
So, rather than enforcing a particular precision, in this work we use $t_{\rm{int}}$ and $\sigma_\mathrm{samp}$ to investigate which parameters are difficult to constrain and gain some physical insight in the problem.

\subsection{Convergence of individual parameters}

\begin{figure}
 \centering
  \includegraphics[width=0.99\columnwidth]{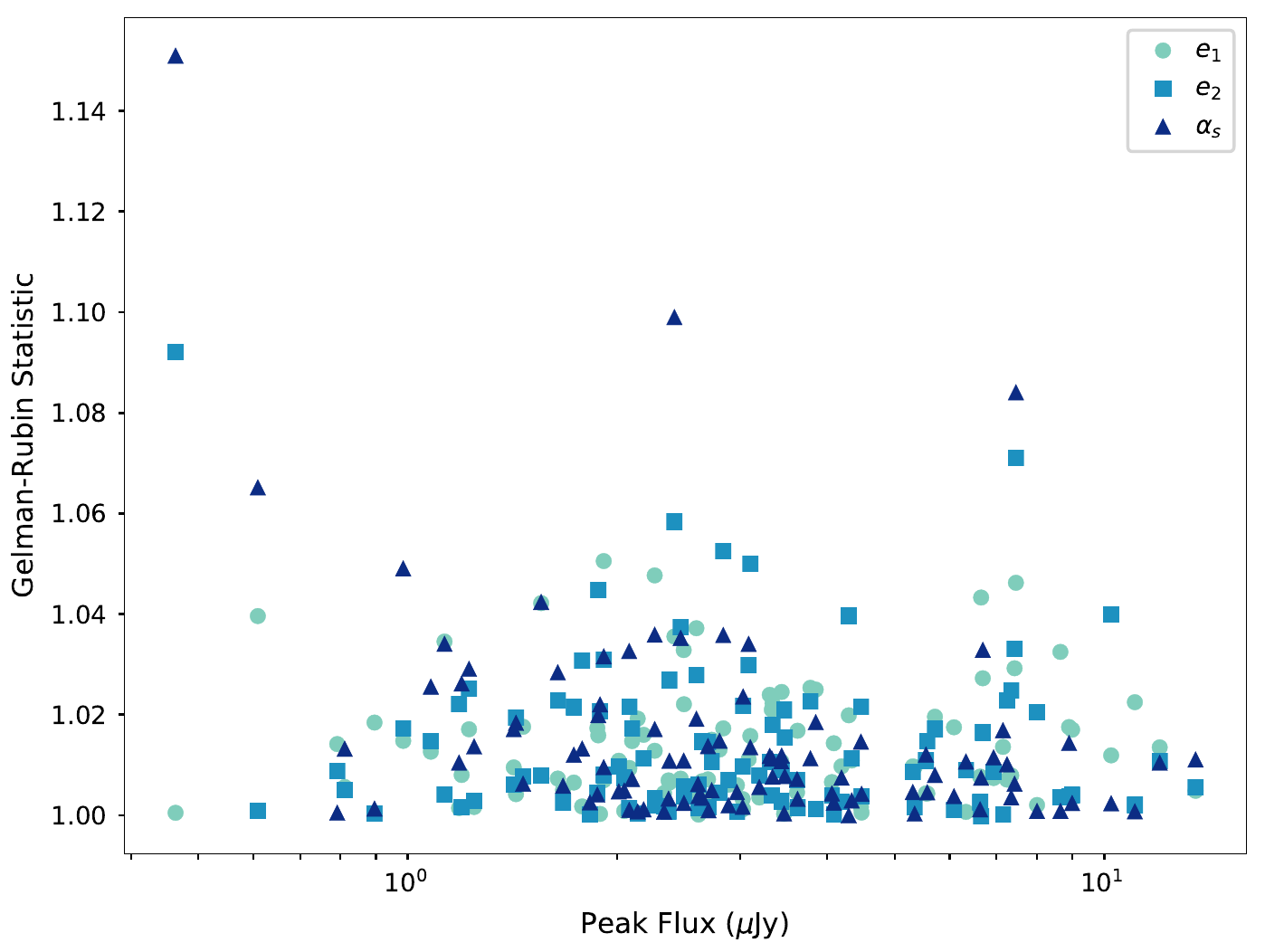} 
  \caption{The Gelman-Rubin statistic as a function of peak flux of the source for all three parameters for the SKA1 100 sources chain.  This indicates the chain is sufficiently converged and also that there may be a loose relationship between how easily a chain converges (i.e. a lower GR value) and peak flux. This relationship is further investigated using the integrated autocorrelation time in Fig.~\ref{fig:iat}.}
  \label{fig:gr}
 \end{figure}

\begin{figure*}
 \centering
 \begin{minipage}{2.1\columnwidth}
 \begin{minipage}{0.99\linewidth}
 \includegraphics[width=0.99\linewidth]{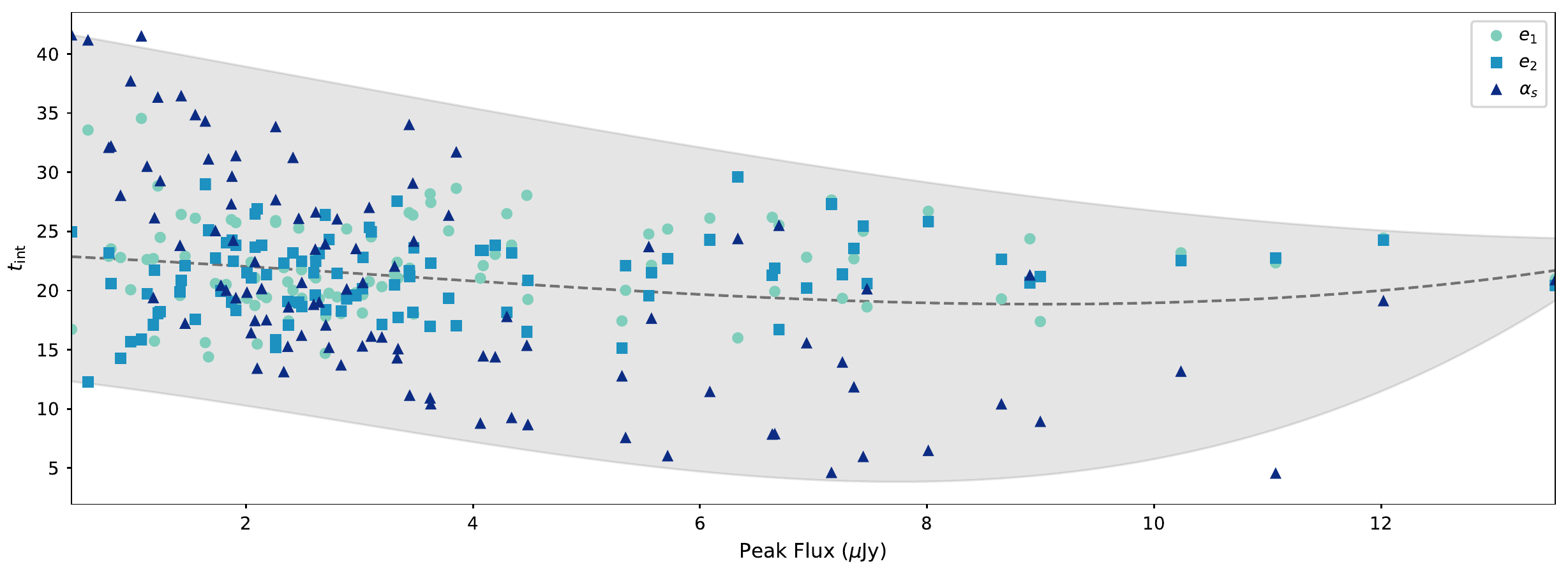} 
 \end{minipage}
 \begin{minipage}{0.99\linewidth}
 \includegraphics[width=0.99\linewidth]{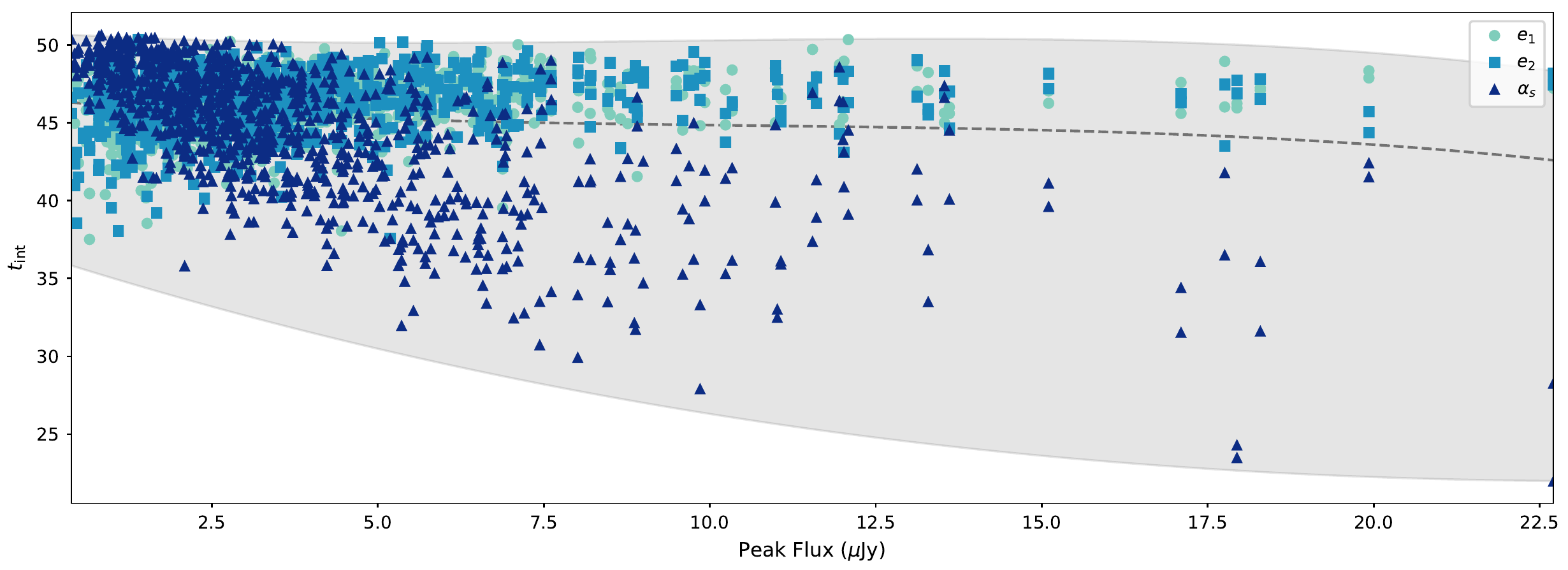} 
 \end{minipage}
 \begin{minipage}{0.99\linewidth}
 \includegraphics[width=0.99\linewidth]{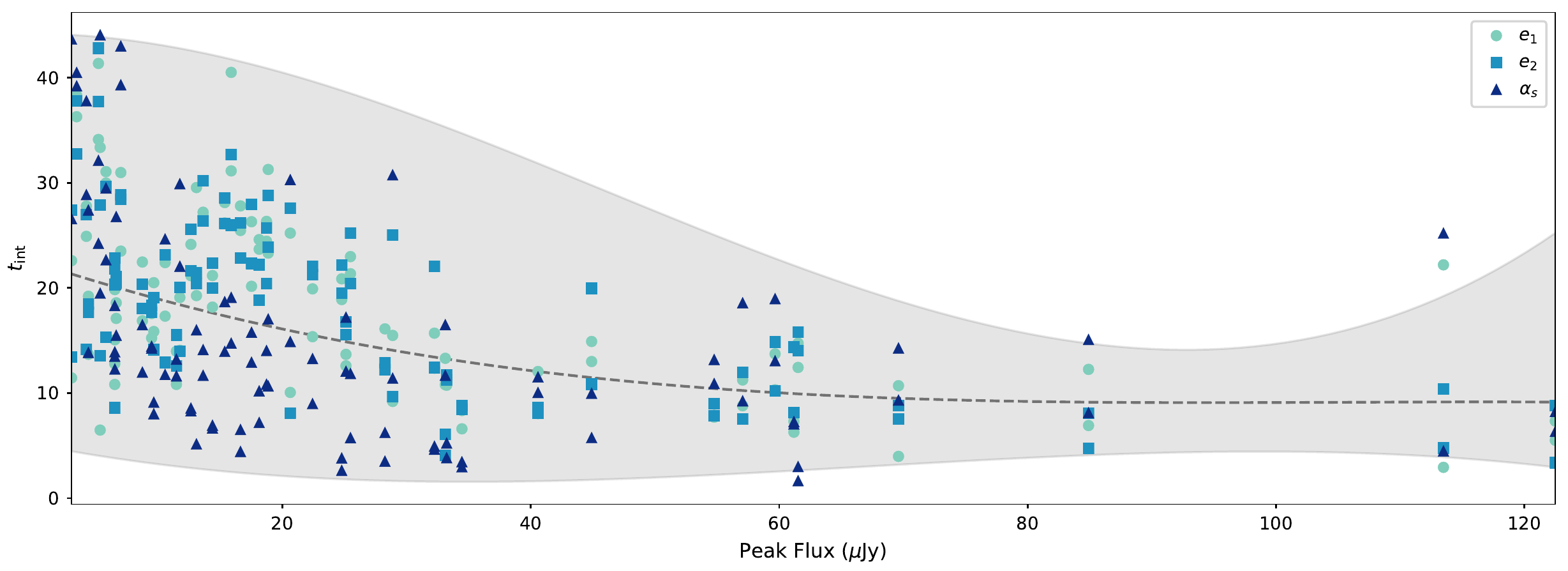} 
 \end{minipage}
 \end{minipage}
 \caption{The integrated autocorrelation time as a function of peak flux of the galaxy for all three parameters, for SKA1 100 sources (top panel), SKA1 1000 sources (middle panel) and SuperCLASS 100 sources (bottom panel). The smaller this value, the more statistically independent samples can be obtained from a chain. In each plot, lines are added to guide the eye that indicate the mean (dashed line) and minimum/ maximum (shaded envelope) in broad bins, interpolated with a cubic spline. For each case, it is clear that the scalelength of brighter sources is easier to constrain, while the effect on ellipticity is more subtle.  The SuperCLASS results indicate that very high SNR sources are significantly easier to constrain.}
\label{fig:iat}
\end{figure*}

\begin{figure*}
 \centering
 \begin{minipage}{2.1\columnwidth}
 \begin{minipage}{0.48\linewidth}
 \includegraphics[width=\linewidth]{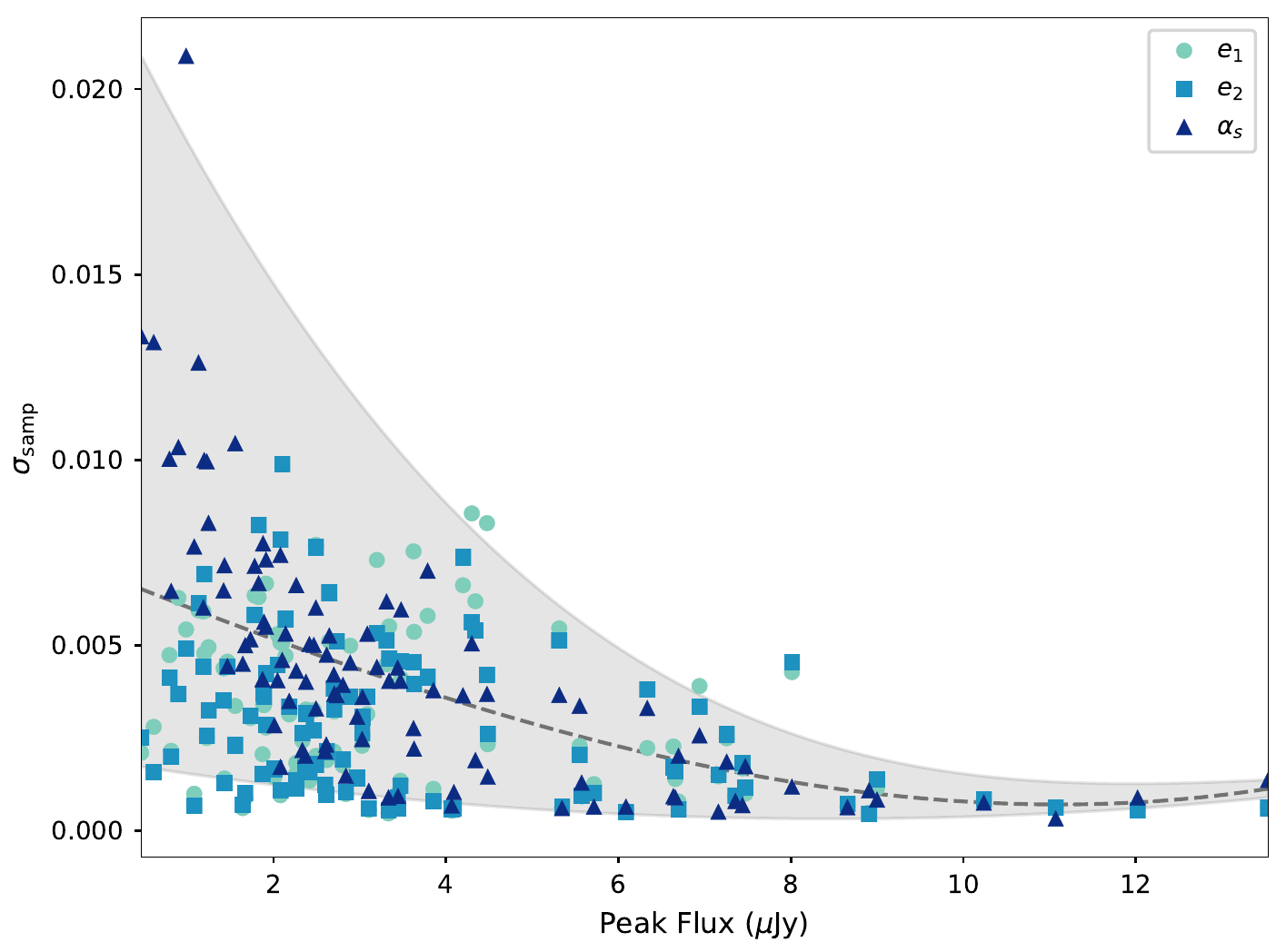} 
 \end{minipage}
 \begin{minipage}{0.48\linewidth}
 \includegraphics[width=\linewidth]{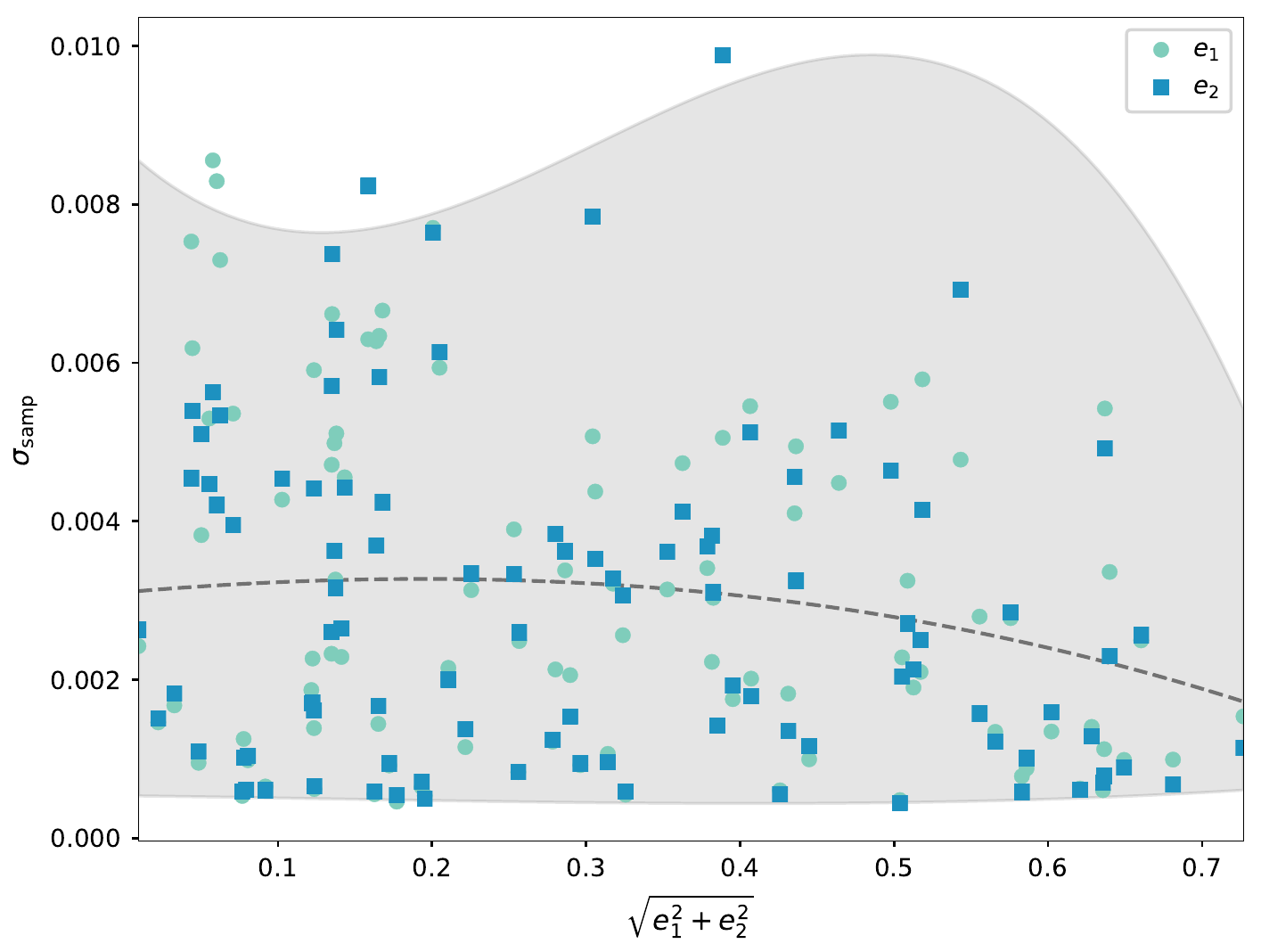} 
 \end{minipage}
 
 \begin{minipage}{0.48\linewidth}
 \includegraphics[width=\linewidth]{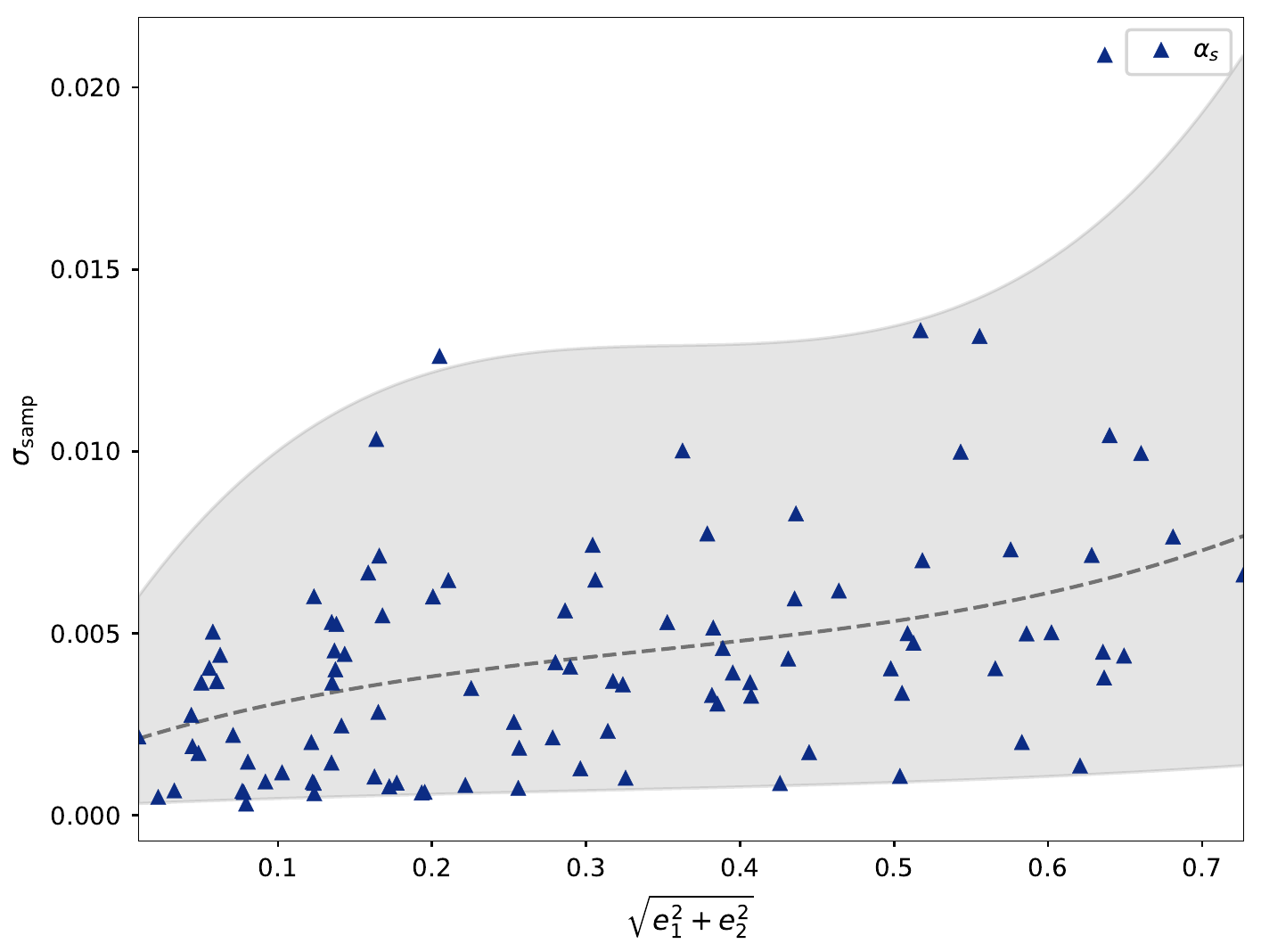} 
 \end{minipage}
 \begin{minipage}{0.48\linewidth}
 \includegraphics[width=\linewidth]{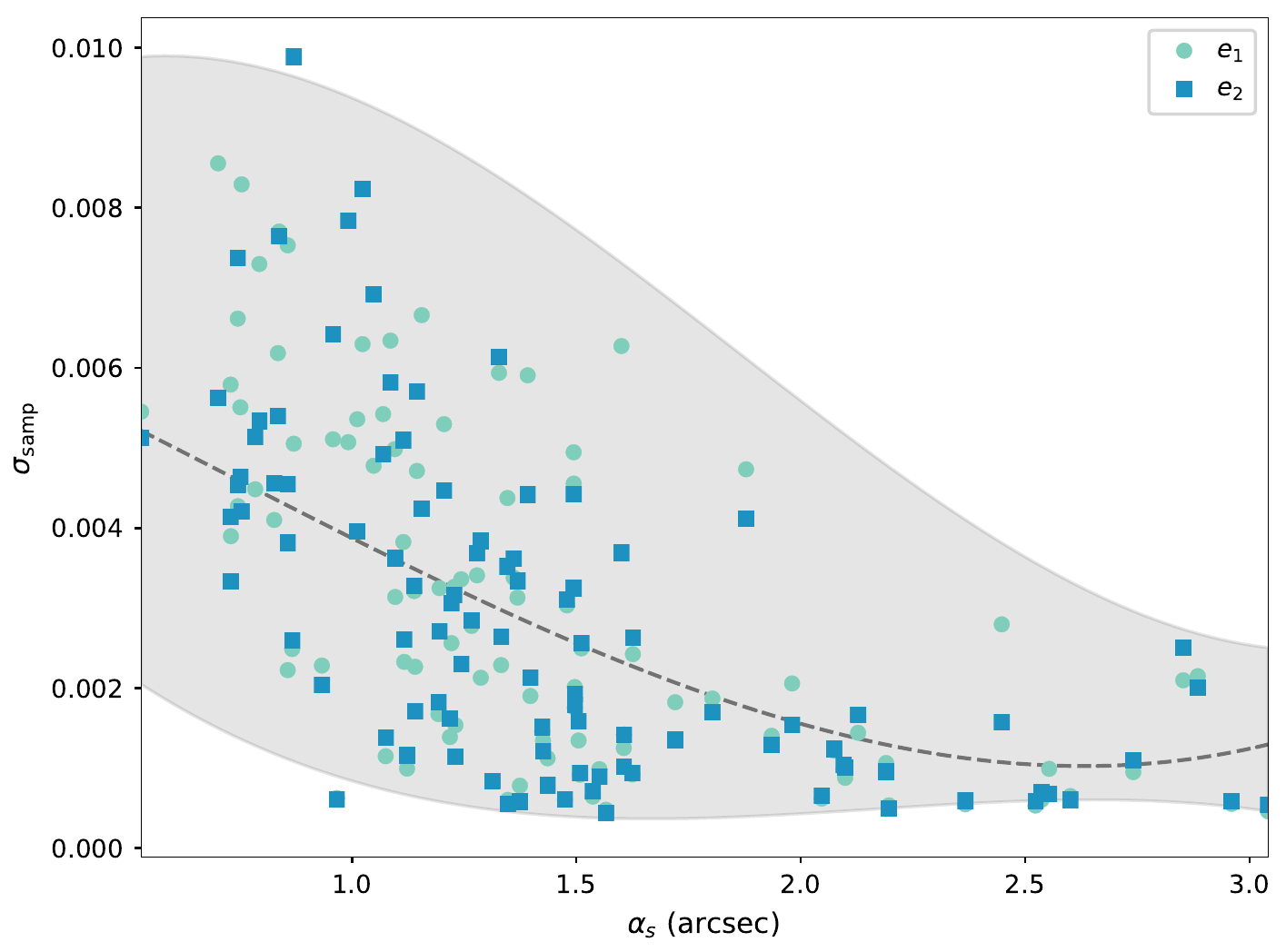} 
 \end{minipage}
 
 \begin{minipage}{0.48\linewidth}
 \includegraphics[width=\linewidth]{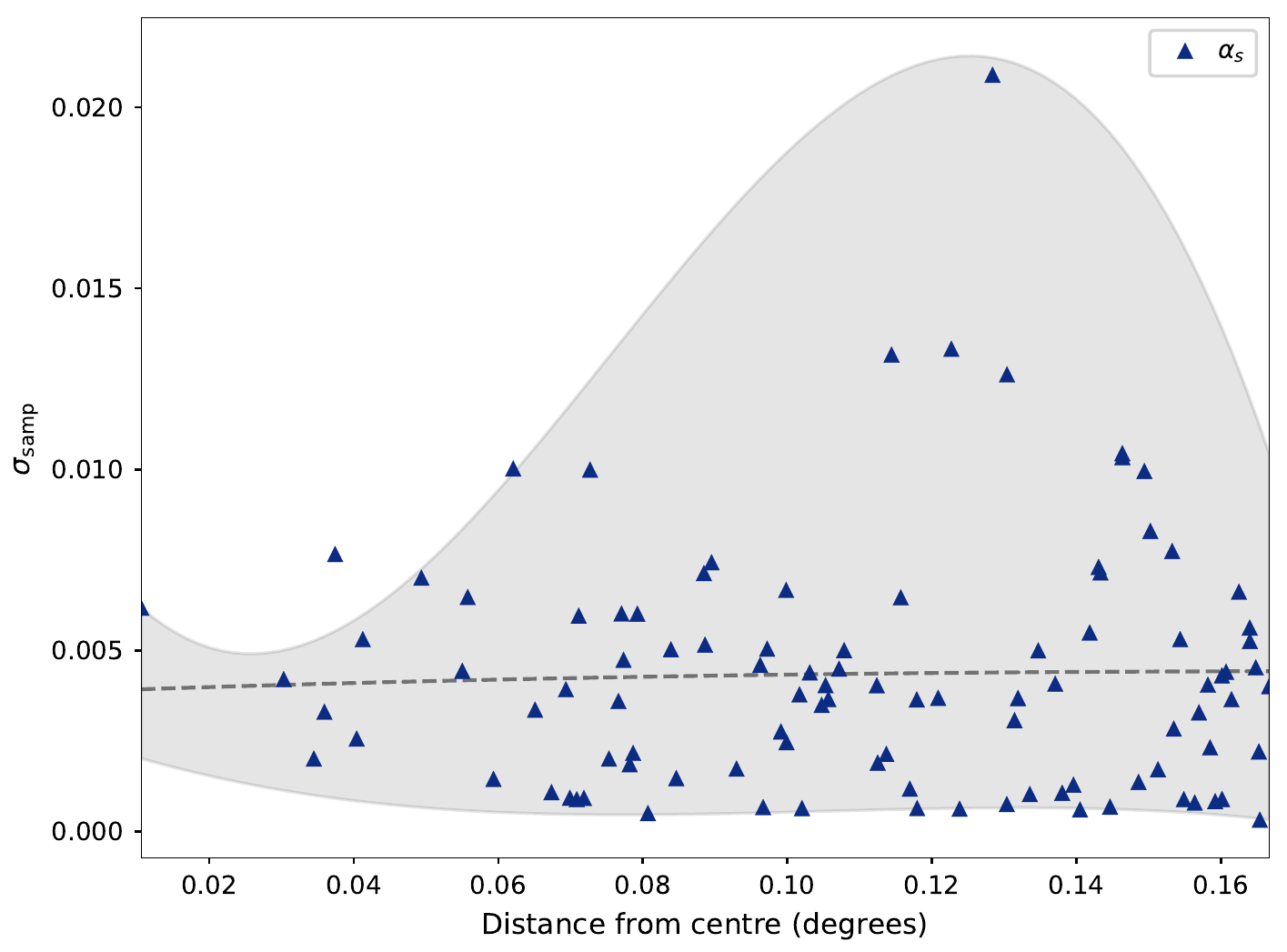} 
 \end{minipage}
 \begin{minipage}{0.48\linewidth}
 \includegraphics[width=\linewidth]{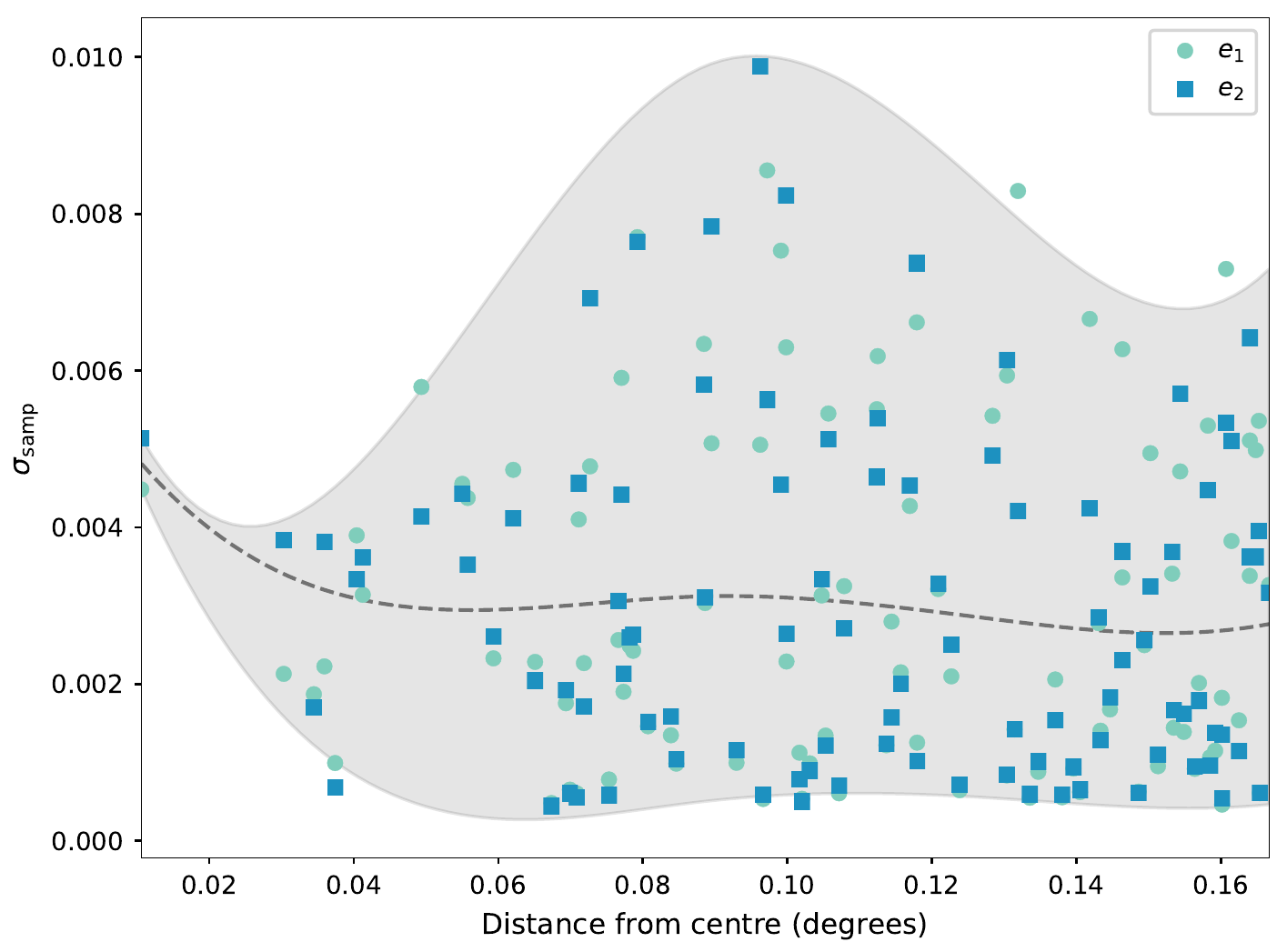} 
 \end{minipage}
 \end{minipage}
 \caption{The sampling uncertainty (that is the contribution to the error on a parameter measurement due to having a chain of finite independent samples) on each of the parameters as a function of different quantities.  \emph{Top left:} All parameters as a function of flux. \emph{Top right:} The ellipticity parameters as a function of the amplitude of ${\bf e}$. \emph{Middle left:}  The scalelength as a function of the amplitude of ${\bf e}$. \emph{Middle right:}  The ellipticity parameters as a function of scalelength. \emph{Bottom left:} The scalelength as a function of the distance of the source from phase centre. \emph{Bottom right:} The ellipticity parameters as a function of the distance of the source from phase centre. In each plot, lines are added to guide the eye that indicate the mean (dashed line) and minimum/ maximum (shaded envelope) in broad bins, interpolated with a cubic spline. All plots use the SKA1 100 sources chain, although the results are similar for the others. The conclusion from this analysis is that the easiest sources to constrain the parameters of tend to be sources that are bright and large. The scalelength is more precisely estimated for round sources, although the corresponding ellipticity uncertainty may be a bit larger. The maximum sampling uncertainty increases with distance from the phase centre, although there is significant scatter.}
\label{fig:convergence}
\end{figure*}

Given that some parameters converge more quickly than others, it is natural to ask what properties about a source make it easy or difficult to constrain its parameters. We investigate this here using the integrated autocorrelation time (see equation~(\ref{eq:tint})) and the sampling uncertainty (equation~(\ref{eq:eff})).

We first test the hypothesis that brighter sources should be easier to constrain. Fig.~\ref{fig:gr} shows the relationship between the Gelman-Rubin statistic and the peak flux for the SKA1 100 sources case. There appears to be some relationship that brighter sources tend to have ``more converged'' parameters but it's difficult to be certain. We investigate this further with the integrated autocorrelation time. As seen in Fig.~\ref{fig:iat}, the integrated autocorrelation time only correlates strongly with flux in the SuperCLASS case where the sources are at a significantly higher signal to noise range. For the SKA1 case, there's a clear indication that the scalelength is easier to constrain for brighter sources, but interestingly there doesn't seem to be a strong relationship between $t_{\rm int}$ and peak flux for the ellipticity parameters. This is not to say the uncertainty on the measured ellipticity parameters doesn't decrease with flux (see Fig.~\ref{fig:results} and Fig.~\ref{fig:SC-100}) but the ellipticity parameters of sources with a higher peak flux are not more easy to measure than faint ones. We do however see a strong correlation with source size in Fig.~\ref{fig:convergence}.

We also consider the sampling uncertainty, which is the uncertainty on the estimate of the mean of a parameter, due to having a finite number of samples in the posterior.  Fig.~\ref{fig:convergence} shows the plots for ``SKA1 100". As expected, the uncertainty decreases with increased flux, even though there is only a subtle relationship between the effective number of samples and flux as shown in Fig.~\ref{fig:iat}. We also find that the sampling error on the ellipticity parameters only has a weak dependence on the actual value of the ellipticity. What is interesting, however, is the relationship between the ellipticity and the scalelength as shown in the middle two panels of Fig.~\ref{fig:convergence}.  The sampling uncertainty of the scalelength increases as the ellipticity does, while the sampling uncertainty for the ellipticity is lower for larger scalelength. For highly elliptical sources, both $e_1$ and $e_2$ need to be determined precisely to get an accurate measure of $\alpha_s$, whereas for more circular sources, this is less important. The middle right panel suggests that for larger sources, the ellipticity is simply easier to measure, as expected since these sources are both easier to resolve and, for a similar peak flux, have a higher integrated flux. The bottom panel indicates an increase in the maximum value of the sampling uncertainty for sources far from the phase centre, which is likely related to a decrease in sensitivity in the beam at the edges of the field of view. 

Finally, within these interesting trends there is a great deal of scatter, implying that the properties of a source alone do not determine how quickly its parameters can be constrained. Because each source interacts with every other source due to the complicated beam pattern of the telescope, it's likely that the exact position of a source relative to the others will also influence constraining power.

\subsection{Computational cost}

\begin{table*}
\centering
 \begin{tabular}{lrcrrrc}
  \hline
  Observation & n. visibilities  & field of view & n. sources & n.samples & n. samples & CPU  time\\
   & & [arcmin$^2$] &  & burn-in & total & per sample [min]\\
  \hline
  SKA1 & 9,266,880 & 400 & 100 & 250 & 15,100 & 0.36 \\
  SKA1 & 9,266,880 & 400 & 1000 & 1000 & 20,900 & 4.64 \\
  SuperCLASS & 345,907,200 & 100 & 100 & 250 & 7,800 & 8.47 \\
  \hline
 \end{tabular}
\caption{A summary of the description of each experiment and the corresponding computational cost to reach convergence (i.e. GR $\approx 1$ for all parameters) in terms of likelihood samples and computational time using 1 CPU core and 2 NVIDIA Tesla K40. ``Burn-in'' refers to the number of samples removed in the initial part of the chain as it finds its way to the best-fitting parameter values. ``n. samples total'' refers to the total length of the chain before burn-in is removed.}
\label{tab:convergence}
\end{table*}

Table \ref{tab:convergence} shows, for all the tests presented in the paper, the number of samples taken before a chain is converged, that naturally increases as more sources are added to the problem. It also shows the average CPU time for computing a chain sample on a single core of Intel Xeon E5-2650 exploiting 2 NVIDIA Tesla K40 GPUs. Obviously the computational time increases both with the number of sources and visibilities.
Note that most of the computational time may be wasted in data transfer between CPU and GPU. Since GPU memory size is much smaller compared to the CPU one, large datasets as the ones produced by radio observations must be split in chunks to be sent through PCIe connection to the available GPUs. Montblanc implements this trying to overlap GPU computation with data transfer, however the current serial version of the code is not scalable with the number of GPUs used because of the many data chunks and transfer bandwidth bottleneck. The new generation of NVIDIA Tesla Pascal GPUs should overcome this issue by exploiting the new NVLink\footnote{https://www.nvidia.com/en-us/data-center/nvlink/} interconnection which maximizes the throughput of multi-GPUs and GPU/CPU system configurations through a larger bandwidth, more links and a better scalability. Moreover, a parallel version of the code is in preparation, implementing the Message Passing Interface paradigm (MPI\footnote{http://www.mpi-forum.org/}), that will also allow the distribution of data and computation among more CPUs of hybrid multi-node architectures. This implementation will be required when dealing with full SKA size datasets although their size may be reduced by working with gridded visibilities, as long as they had been gridded with an appropriate set of gridding kernels \citep{SKA-ECP}. 

\section{Discussion on AGN contamination}
\label{AGN}
The continuum faint radio sky observed at 1-2 GHz for weak lensing surveys is dominated by SF galaxy populations, however a non-negligible fraction of sources is expected to be associated with Active Galactic Nuclei (AGN) \citep{Jarvis15, COSMOS2017, GOODSN2018, TRECS}. Therefore an investigation about AGN classification and visibility shape modelling should be performed in order to handle their contamination. 
In deep radio fields two AGN populations are detected: radio-loud (RL), where synchrotron emission is dominated by large-scale relativistic jets and the lobes that the jets inflate, and radio-quiet (RQ), not showing jet-related emission and featuring much weaker radio emission.
RL AGNs are well known to dominate the bright portion of the radio counts above 0.5 mJy at 1.4 GHz, but moving towards fluxes below few hundreds $\mu$Jy this population should be progressively outnumbered by RQ AGN \citep{Mancuso2017}.

In \cite{Chang04} a first attempt to model the shape of AGN components in the visibility domain is applied to the VLA FIRST survey where the dominating AGN population is RL due to the low sensitivity of the radio survey. They used shapelets to fit simultaneously all sources in a given pointing, including AGN lobes, thus not contaminating fainter sources in the primary beam and enabling cosmic shear measurement.
However whilst AGN will indeed be lensed along with SF galaxies, their complicated morphologies mean noise due to intrinsic shape dispersion will be large, and fitting simple parameterised profiles will lead to large model biases. Hence we expect to have to remove RL AGN to get more accurate results. 
Several classification methods, mainly based on the comparison with other wave-bands catalogs of the same observed area, are already available and/or under further investigation, e.g. \cite{Padovani2016, Barger2017}. Such classification schemes will never be completely perfect, and leakage of RL AGN into the weak lensing sample will potentially create residual model biases and increase in shape noise. Initial studies indicate that successful classification rates of $\sim90\%$ should be sufficient \citep{mccallum}.

Deep SKA precursor surveys such as e-MERGE\footnote{http://www.e-merlin.ac.uk/legacy/projects/emerge.html} highlight that in the weak lensing regime a sizeable fraction (20-30\%) of SF galaxies may host RQ AGN \citep{Bonzini2013, Delvecchio2017, GOODSN2018}, thus particular interest should be given to these systems. Observed data, e.g. \citet{Guidetti2017}, suggest that radio emission from RQ AGN is compact. Therefore they may be modelled by a combination of exponential discs and Gaussians.

\section{Conclusions}
\label{conclusions}
We have presented a new Bayesian method for measuring SF galaxy shape parameters from visibility data of radio observations. The method extends the BIRO technique by implementing a joint model fitting of the ellipticity and scalelength of all exponential sources in the field of view. It applies a Hamiltonian Monte Carlo technique, and can be easily extended for the measurement of other galaxy parameters such as position, flux and in band spectral index. Since we follow the RIME approach (see Section~\ref{sec:likelihood}) for computing the model visibilities, it can also be extended for a simultaneous inference of scientific and instrumental parameters  \citep{BIRO}. 

We tested the method on simulated observations of up to 1000 galaxies adopting the SKA-MID Phase~1 UV coverage at 1.4~GHz. We were able to recover with a good accuracy the original galaxy shape parameters (see Table~\ref{tab:SKAbest-fit}) at the source density of the proposed SKA1 radio weak lensing survey (2.7~gal/arcim$^2$) with SNR$_\mathrm{vis} \ge 10$. As expected, the joint fitting approach improves measurements of galaxy ellipticies obtained with the \textit{RadioLensfit} method at the same source density because it removes the neighbour bias introduced by the source extraction procedure. On the other hand, HMC has a long computation time, as shown in Table~\ref{tab:convergence}, which can be reduced with the new GPUs architectures and a higher level of code parallelisation exploiting a hybrid multi-node HPC system.

We also applied this method to the simulation of a fraction (100~arcmin$^2$) of SuperCLASS, a precursor radio weak lensing survey performed combining observations of e-MERLIN and JVLA radio telescopes (although again we caution not to directly compare results from the simplified SKA simulation and realistic SuperCLASS simulation). Since the assumed minimum 10$\sigma$ detection in the image domain corresponds to a SNR$_\mathrm{vis} \ge 18$, we obtained a faster convergence of the chains and better fitting than with SKA1 experiments. In fact, the convergence analysis showed that the parameters of bright sources are easier to constrain. It also showed that ellipticity sampling uncertainty is strongly correlated with the source size and it is lower for large sources. 

These results show that working with visibilities may provide a more accurate source characterization, and more reliable and complete source catalogs, potentially offering a novel measurement approach for future SKA surveys. In this case, more computing resources may be required in order to deal with the large volume of data that will be produced by the future surveys and the complexity of the data analysis algorithms. Moreover, further investigation on AGN shape modelling should be performed as real observations will also contain a non-negligible fraction of AGN population.

\section*{Acknowledgements}
We thank Simon Perkins for support with Montblanc, Isabella Prandoni for useful discussions about AGN and e-MERLIN and JVLA extended source detection. We also thank the SuperCLASS collaboration for access to and assistance with their simulations.

The financial assistance of the National Research Foundation (NRF) towards this research is hereby acknowledged. Opinions expressed and conclusions arrived at, are those of the authors and are not necessarily to be attributed to the NRF. MR acknowledges the support of the Science and Technology Facilities Council via an SKA grant. FBA acknowledges the support of the Royal Society via a Royal Society URF award.

\bibliographystyle{mn2e_trunc8}
\bibliography{master}

\appendix
\section{Likelihood Gradient} 
\label{appendix}

The Montblanc code returns the chi-squared value computed comparing data visibilities~$\tilde V_i$ with the sky model visibilities~$V_i({\bf x})$:
\begin{equation}
\chi^2({\bf x})=\sum_i \frac{|\tilde V_i - V_i({\bf x})|^2}{\sigma_i^2}.
\end{equation}
From it the likelihood can be easily obtained as $\mathcal{L}({\bf x}) \propto \exp \left[- \chi^2({\bf x})/2\right]$. 
Similarly we can compute the likelihood gradient with respect to a set of parameters $p_1, \ldots, p_M$ by adding to Montblanc the computation of the corresponding chi-squared partial derivatives:
\begin{eqnarray}
\label{gradient}
\frac{\partial \chi^2({\bf x})}{\partial p_j}  & = & -2 \sum_i \frac{\Re \tilde V_i - \Re V_i({\bf x})}{\sigma_i^2}\cdot \frac{\partial \Re V_i({\bf x})}{\partial p_j} + \nonumber \\
&&  -2 \sum_i \frac{\Im \tilde V_i - \Im V_i({\bf x})}{\sigma_i^2} \cdot \frac{\partial \Im V_i({\bf x})}{\partial p_j},
\end{eqnarray}
where $\Re$ and $\Im$ denote the real and imaginary part of the complex visibilities.
Adopting visibility formulation as in equation~(\ref{eq:vis}), galaxy shape parameters $p_{s,k}$ are contained only in the brightness matrix of source $s$, therefore 
\begin{equation}
\frac{\partial V_{tpq\lambda}}{\partial p_{s,k}} = \frac{\partial V_{tpq\lambda s}}{\partial p_{s,k}} = K_{tps\lambda} \frac {\partial B_{s\lambda}}{\partial p_{s,k}} K^H_{tqs\lambda}.  
\end{equation}
From equation~(\ref{eq:brightness}), we have the following partial derivatives of the model visibilities with respect to S\'{e}rsic shape parameters $\alpha_s, {\bf e}_s=(e_{1,s}, e_{2,s})$ of each galaxy $s=1, \ldots, N$ in the field of view: 
\begin{eqnarray}
\frac{\partial V_{tpq\lambda}}{\partial \alpha_s} & = & - \frac{12 \pi^2\alpha_s |\mathbf{A}_s^{-T}\mathbf{k}|^2}{1+4\pi^2 \alpha^2_s |\mathbf{A}_s^{-T}\mathbf{k}|^2} \cdot V_{tpq\lambda s}, \\
\frac{\partial V_{tpq\lambda}}{\partial e_{1,s}} & = & - \frac{12 \pi^2\alpha_s^2 f_1({\bf e}_s)}{1+4\pi^2 \alpha^2_s |\mathbf{A}_s^{-T}\mathbf{k}|^2} \cdot V_{tpq\lambda s}, \\
\frac{\partial V_{tpq\lambda}}{\partial e_{2,s}} & = & - \frac{12 \pi^2\alpha_s^2 f_2({\bf e}_s)}{1+4\pi^2 \alpha^2_s |\mathbf{A}_s^{-T}\mathbf{k}|^2} \cdot V_{tpq\lambda s},
\end{eqnarray}
with
\begin{eqnarray}
f_1({\bf e}_s) = \frac{\left[ (1+e_{1,s})u + e_{2,s}v\right] u}{(1-e_{1,s}^2 - e_{2,s}^2)^2}+\frac{2e_{1,s}\left[ (1+e_{1,s})u + e_{2,s}v\right]^2 }{(1-e_{1,s}^2 - e_{2,s}^2)^3}  \nonumber \\
- \frac{ \left[ e_{2,s}u + (1-e_{1,s})v\right]v}{(1-e_{1,s}^2 - e_{2,s}^2)^2} +\frac{2e_{1,s} \left[ e_{2,s}u + (1-e_{1,s})v\right] ^2 }{(1-e_{1,s}^2 - e_{2,s}^2)^3} \nonumber
\end{eqnarray}  
and
\begin{eqnarray}
f_2({\bf e}_s) =  \frac{\left[ (1+e_{1,s})u + e_{2,s}v\right] v}{(1-e_{1,s}^2 - e_{2,s}^2)^2}+\frac{2e_{2,s}\left[ (1+e_{1,s})u + e_{2,s}v\right]^2 }{(1-e_{1,s}^2 - e_{2,s}^2)^3}  \nonumber \\
+ \frac{ \left[ e_{2,s}u + (1-e_{1,s})v\right]u}{(1-e_{1,s}^2 - e_{2,s}^2)^2} +\frac{2e_{2,s} \left[ e_{2,s}u + (1-e_{1,s})v\right] ^2 }{(1-e_{1,s}^2 - e_{2,s}^2)^3}. \nonumber
\end{eqnarray}

\end{document}